\UseRawInputEncoding

\documentclass[a4paper,fleqn]{cas-dc}

\usepackage[numbers]{natbib}

\usepackage{amssymb}
\usepackage{amsmath}
\usepackage{soul}
\usepackage{listings}
\usepackage{unitsdef}

\definecolor{commentsColor}{rgb}{0, 0.6, 0}
\definecolor{keywordColor}{rgb}{0, 0, 1.0}

\lstdefinelanguage{Python3}[]{Python}{
    morekeywords={await, async},
    deletekeywords={len}
}

\DeclareMathOperator*{\argmin}{\arg\!\min}
 
\let\oldhat\hat

\renewcommand{\hat}[1]{\skew{7}\oldhat{#1}}

\makeatletter 
\newcommand\semiHuge{\@setfontsize\semiHuge{20}{27.38}}
\makeatother

\begin{document}
\let\WriteBookmarks\relax
\def\floatpagepagefraction{1}
\def\textpagefraction{.001}

\shorttitle{}

\shortauthors{C. Cueto et~al.}

\title [mode = title]{Stride: a flexible software platform for high-performance ultrasound computed tomography}


%
\author[1]{Carlos Cueto}

\cormark[1]

\ead{c.cueto@imperial.ac.uk}

\affiliation[1]{
    organization={Department of Bioengineering},
    addressline={Imperial College London}, 
    city={London},
    postcode={SW7 2AZ}, 
    country={United Kingdom}
}

\author[1]{Oscar Bates}

\author[2]{George Strong}

\affiliation[2]{
    organization={Department of Earth Science and Engineering},
    addressline={Imperial College London}, 
    city={London},
    postcode={SW7 2AZ}, 
    country={United Kingdom}
}

\author[2]{Javier Cudeiro}

\author[3]{Fabio Luporini}

\affiliation[3]{
    organization={Devito Codes},
    addressline={}, 
    city={London},
    postcode={}, 
    country={United Kingdom}
}

\author[2]{\`{O}scar Calder\'{o}n Agudo}

\author[2]{Gerard Gorman}

\author[2]{Lluis Guasch}

\cormark[2]

\ead{l.guasch08@imperial.ac.uk}

\author[1]{Meng-Xing Tang}

\cormark[2]

\ead{mengxing.tang@imperial.ac.uk}

\cortext[cor1]{Corresponding author}
\cortext[cor2]{Principal corresponding author}

\begin{abstract}
\noindent
\noindent
\textit{Background and objective:} Advanced ultrasound computed tomography techniques like full-waveform inversion are mathematically complex and orders of magnitude more computationally expensive than conventional ultrasound imaging methods. This computational and algorithmic complexity, and a lack of open-source libraries in this field, represent a barrier preventing the generalised adoption of these techniques, slowing the pace of research, and hindering reproducibility. Consequently, we have developed Stride, an open-source Python library for the solution of large-scale ultrasound tomography problems. \\

\noindent
\textit{Methods:} On one hand, Stride provides high-level interfaces and tools for expressing the types of optimisation problems encountered in medical ultrasound tomography. On the other, these high-level abstractions seamlessly integrate with high-performance wave-equation solvers and with scalable parallelisation routines. The wave-equation solvers are generated automatically using Devito, a domain-specific language, and the parallelisation routines are provided through the custom actor-based library Mosaic. \\

\noindent
\textit{Results:} We demonstrate the modelling accuracy achieved by our wave-equation solvers through a comparison (1) with analytical solutions for a homogeneous medium, and (2) with state-of-the-art modelling software applied to a high-contrast, complex skull section. Additionally, we show through a series of examples how Stride can handle realistic numerical and experimental tomographic problems, in 2D and 3D, and how it can scale robustly from a local multi-processing environment to a multi-node high-performance cluster. \\

\noindent
\textit{Conclusions:} Stride enables researchers to rapidly and intuitively develop new imaging algorithms and to explore novel physics without sacrificing performance and scalability. This will lead to faster scientific progress in this field and will significantly ease clinical translation.

\end{abstract}

\ExplSyntaxOn
\keys_set:nn { stm / mktitle } { nologo }
\ExplSyntaxOff
\maketitle

\section{Introduction}
\label{sec:intro}

Ultrasound computed tomography techniques such as full-waveform inversion (FWI) have the potential to produce high-resolution, 3D reconstructions of tissues such as the breast \cite{Wiskin20173-DResults,Sandhu2015FrequencyTransducer}, the limbs \cite{Wiskin2020FullContrast}, or the adult human brain \cite{Guasch2020Full-waveformBrain}. However, generalised adoption of these techniques is hindered by the fact that tomography algorithms are computationally demanding and algorithmically complex, while existing medical tomography codes are, as far as we are aware, closed source, difficult to maintain, and slow to adapt to new research.

FWI is a technique, originally developed in the field of geophysics, that produces reconstructions of tissue properties by solving an associated inverse problem. FWI is computationally expensive because, for realistic 3D problems, it requires the solution of thousands of partial-differential equations (PDEs) and the storage of hundreds of gigabytes of memory at every iteration in order to estimate billions of parameters. At the same time, FWI is algorithmically challenging due to the non-linear, non-convex nature of the inverse problem being solved. Therefore, any software for solving FWI problems has to address its computational and algorithmic needs, but should also emphasise the high-level, problem-specific abstractions that are necessary to ease the adoption of these tomographic techniques.

In the fields of geophysics and seismic exploration, different approaches have been taken by open-source libraries to solve these issues. On one hand, libraries like Madagascar \cite{Fomel2013Madagascar:Experiments}, SimPEG \cite{Cockett2015SimPEG:Applications} and PySIT \cite{HewettPySIT:Toolbox}, have managed to provide flexibility and high-level abstractions, but have done so at the expense of performance. On the other hand, libraries like SAVA \cite{KoehnSAVA:Media} and JavaSeis \cite{Hassanzadeh1997JavaSeis:Services} have focused on performance at the expense of flexibility and extensibility. As a way to bridge the gap between these two extremes, libraries such as SeisFlows and Pyatoa \cite{Modrak2018SeisFlowsFlexibleSoftware,Chow2020AnZealand}, jInv \cite{Ruthotto2017JInv--aEstimation}, and Waveform \cite{DaSilva2019AProblems} provide flexible interfaces in high-abstraction languages like Python, Julia or MATLAB that interface with high-performance, hand-tuned solvers. LASIF and Inversionson \cite{Krischer2015Large-scaleFramework,Thrastarson2021Inversionson:Inversions} have followed a similar approach by providing modular seismic work-flow management that wraps a high-performance tomography solver.

\begin{figure*}[t]
    \centering
    \includegraphics[width=17.5cm]{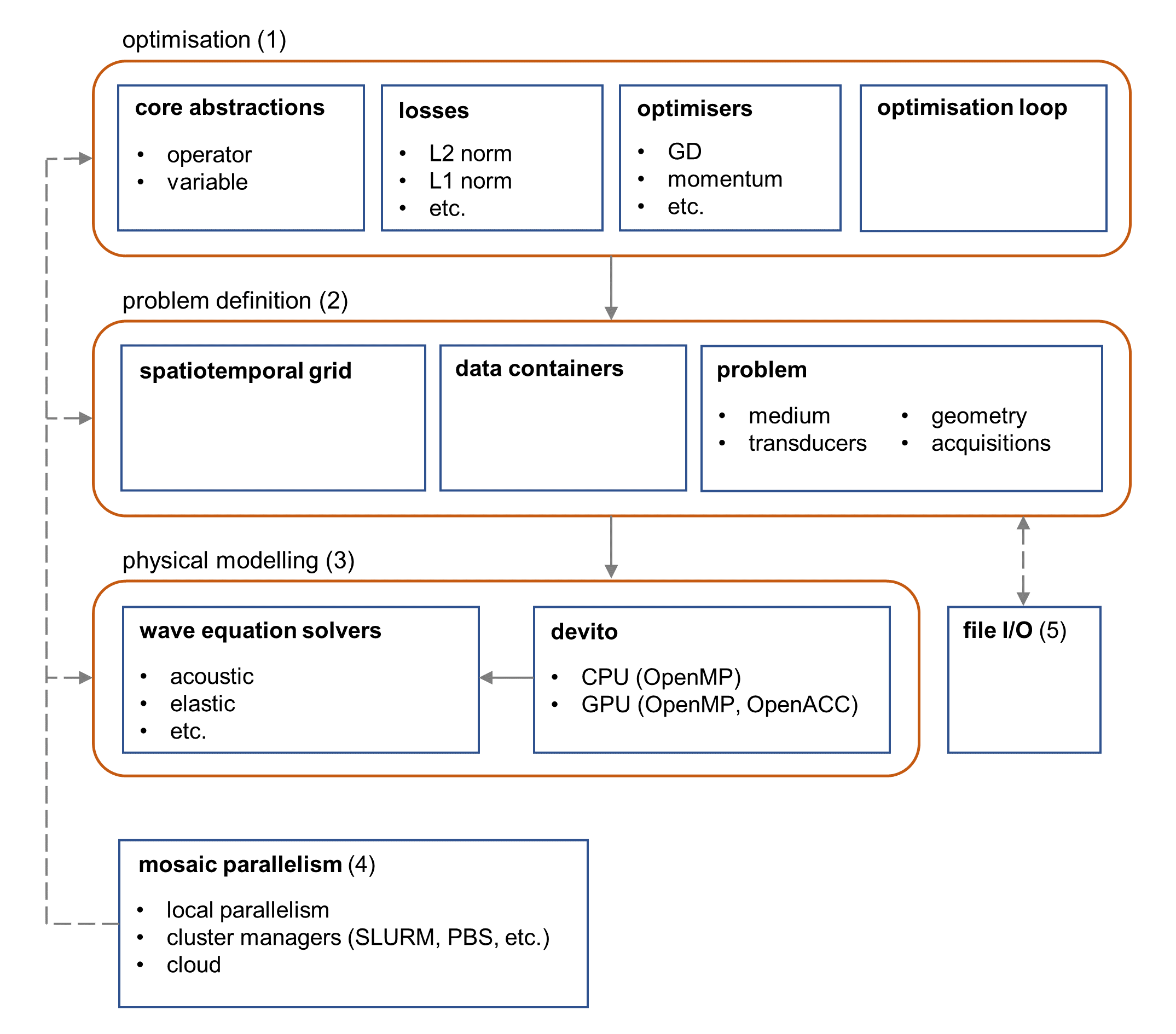}
    \caption{Schematic representation of the Stride software structure. A series of basic abstractions for solving optimisation problems are provided (1), based on which the tomographic problem is expressed (2). The tomographic problem becomes fully defined when appropriate physical modelling is introduced (3). The execution of Stride is parallelised using the custom library Mosaic (4), and tools are provided to save and load its details (5).}
    \label{fig:schematic}
\end{figure*}

Recently JUDI \cite{Witte2019AJulia}, written in Julia, has gone a step further by providing high-level abstractions in a modern language together with high-performance solvers that are automatically generated by the domain-specific language (DSL) Devito \cite{Louboutin2019DevitoExploration, Luporini2020ArchitectureComputation}. Automatic code generation for solvers is increasingly important with an ever growing number of specialised architectures, from traditional central processing units (CPUs) to graphical processing units (GPUs) and field-programmable gate arrays (FPGAs), as well as associated parallel programming languages (Cuda, OpenACC, etc.). Fine tuning codes for each of them by hand would be a daunting task for most researchers, whereas DSLs like Devito can generate code that is automatically tuned for each target architecture and parallel language. In doing so, DSLs also increase productivity by simplifying the implementation of new types of physics and discretisations.

The high computational complexity of FWI also requires, for realistic problems, that codes can be deployed to specialised high-performance computing (HPC) systems like multi-node clusters or cloud computing services. This represents a further barrier for domain scientists, who are generally not proficient in the use of HPC systems. Of the reviewed geophysical and seismic libraries, only some of them, such as LASIF and Inversionson, and SeisFlows and Pyatoa, have been designed with HPC deployment and scaling in mind.

Here, we present Stride, an open-source Python library for medical ultrasound tomography that emphasises flexibility and modularity, high performance, and scalability. It achieves this, firstly, through high-level, domain-specific abstractions and heuristics. Secondly, by integrating with the automatic code generation library Devito. Finally, we introduce a parallelisation library for seamless HPC deployment and scaling. Stride is available on GitHub\footnote{https://github.com/trustimaging/stride} and a complete documentation of its interfaces is available online\footnote{https://stridecodes.readthedocs.io}.

The remaining of this paper is structured as follows: in Sec. \ref{sec:methods}, we will present an overview of the structure of Stride, followed by a more detailed exploration of each of its components with accompanying examples; in Sec. \ref{sec:results}, we will assess the accuracy of the wave-equation solvers provided by Stride, and we will present examples of tomographic reconstructions in 2D and 3D, using both numerical and experimental data; finally, we will present our discussion and proceed to our conclusions in Sec. \ref{sec:discussion} and Sec. \ref{sec:conclusions}, respectively.

\section{Methods}
\label{sec:methods}

\subsection{Software structure}

Stride has been designed to address the computational and algorithmic complexity of tomographic imaging by providing high-level interfaces that are modular and extensible, and that closely match the mental framework of domain specialists. It has been implemented in Python, a high-level, interpreted programming language that provides characteristics such as portability, ease of use, and dynamic typing. We have chosen Python because it is the \textit{de facto} language for scientific computing and machine learning, with a large community and package ecosystem.

The high-level interfaces provided by Stride are aimed at addressing five fundamental aspects in high-performance ultrasound computed tomography (Fig. \ref{fig:schematic}):

\begin{enumerate}
    \item first, abstractions and tools are provided for the solution of optimisation problems, which are the basis for most tomographic imaging algorithms;
    \item based on these, a series of classes encapsulate the definition of the tomographic problem being solved, e.g. the transducers employed or the signals used to excite them;
    \item the relevant physical processes, such as acoustic or elastic wave propagation, are then modelled by using appropriate solvers that execute high-performance code through DSLs like Devito;
    \item scaling of these algorithms, from a local workstation to HPC clusters, is achieved by using an integrated parallelisation library called Mosaic;
    \item finally, tools are provided for saving and loading the different components of the problem using a standardised file format.
\end{enumerate}

Each of these will be presented in detail in the following five sections.

\subsection{Abstractions for solving optimisation problems}

Techniques such as ultrasound computed tomography, optoacoustic tomography \cite{Arridge2016OnTomography}, or even ultrasound calibration techniques like spatial response identification \cite{Cueto2021SpatialImaging,Cueto2022SpatialTomography}, are commonly formulated as mathematical optimisation problems, which are solved numerically by using local methods like gradient descent. Therefore, a fundamental necessity when implementing these techniques is the availability of abstractions that allow us to pose our optimisation problems, calculate gradients of those problems with respect to the relevant parameters, and then apply these gradients through some local optimisation algorithm. In the next paragraphs, we introduce the abstractions that, being at the core of Stride, enable the solution of such inverse problems.

Consider a continuously differentiable function $f(\mathbf{y})$, which can be expressed as $f(\mathbf{y}) = \left\langle \hat{f}(\mathbf{y}), 1 \right\rangle$ with some adequate function $\hat{f}(\mathbf{y})$ and some bilinear form $\left\langle \alpha, \beta \right\rangle$. We know that the derivative of $f(\mathbf{y})$ with respect to $\mathbf{y}$ is,

\begin{equation}
    \nabla_\mathbf{y} f(\mathbf{y}) \delta\mathbf{y} 
    = \left\langle \nabla_\mathbf{y} \hat{f}(\mathbf{y}) \delta\mathbf{y}, 1 \right\rangle
    = \left\langle \nabla_\mathbf{y} \hat{f}(\mathbf{y}), \delta\mathbf{y} \right\rangle
\end{equation}

\noindent
where $\nabla_\mathbf{y} f(\mathbf{y})\delta\mathbf{y}$ represents the derivative of an operator $f(\mathbf{y})$ in the direction $\delta\mathbf{y}$, and the derivative is by definition linear in the differentiation direction. Consider now that $\mathbf{y} = \mathbf{g}(\mathbf{z})$ is another continuously differentiable function. Then the derivative of $f(\mathbf{y})$ with respect to $\mathbf{z}$ is,

\begin{equation}
    \nabla_\mathbf{z} f(\mathbf{y}) \delta\mathbf{z} 
    = \left\langle \nabla_\mathbf{y} \hat{f}(\mathbf{y}), \delta\mathbf{y} \right\rangle
    = \left\langle \nabla_\mathbf{y} \hat{f}(\mathbf{y}), \nabla_\mathbf{z} \mathbf{g}(\mathbf{z}) \delta\mathbf{z} \right\rangle
\end{equation}

\noindent
by virtue of the product rule. At this point, we introduce the concept of the adjoint of an operator. Given an operator $D\cdot$, its adjoint is $D^*\cdot$, defined so that $\left\langle a, Db  \right\rangle = \left\langle b, D^*a  \right\rangle$. Then, we can rewrite the expression as,

\begin{equation}
\begin{split}
    \nabla_\mathbf{z} f(\mathbf{y}) \delta\mathbf{z} 
    &= \left\langle \nabla_\mathbf{y} \hat{f}(\mathbf{y}), \nabla_\mathbf{z} \mathbf{g}(\mathbf{z}) \delta\mathbf{z} \right\rangle \\
    &= \left\langle \nabla_\mathbf{z}^* \mathbf{g}(\mathbf{z}) \nabla_\mathbf{y} \hat{f}(\mathbf{y}), \delta\mathbf{z} \right\rangle
\end{split}
\end{equation}

That is, the derivative of function $f(\mathbf{y})$ with respect to $\mathbf{z}$ can be calculated by finding the derivative of $\hat{f}(\mathbf{y})$ with respect to its input $\mathbf{y}$ and then applying the adjoint of the Jacobian of $\mathbf{g}(\mathbf{z})$ on the result. In the discrete case, this is equivalent to the Jacobian-vector product. 

Similarly, if we added a third function $\mathbf{z} = \mathbf{h}(\mathbf{x})$, then the same result could be obtained for the derivative of $f(\mathbf{y})$ with respect to $\mathbf{x}$,

\begin{equation}
\begin{split}
    \nabla_\mathbf{x} f(\mathbf{y}) \delta\mathbf{x} 
    &= \left\langle \nabla_\mathbf{z}^* \mathbf{g}(\mathbf{z}) \nabla_\mathbf{y} \hat{f}(\mathbf{y}), \delta\mathbf{z} \right\rangle \\
    &= \left\langle \nabla_\mathbf{z}^* \mathbf{g}(\mathbf{z}) \nabla_\mathbf{y} \hat{f}(\mathbf{y}), \nabla_\mathbf{x} \mathbf{h}(\mathbf{x}) \delta\mathbf{x} \right\rangle \\
    &= \left\langle \nabla_\mathbf{x}^* \mathbf{h}(\mathbf{x}) \nabla_\mathbf{z}^* \mathbf{g}(\mathbf{z}) \nabla_\mathbf{y} \hat{f}(\mathbf{y}), \delta\mathbf{x} \right\rangle
\end{split}
\end{equation}

\noindent
and the same procedure could be followed for any arbitrary chain of functions for whose inputs we wanted to calculate a derivative. This procedure, known as the adjoint method or backpropagation in the field of machine learning, is effectively the reverse mode that automatic differentiation libraries provide to calculate derivatives, albeit in the continuous limit. This is the core abstraction used in Stride.

Stride considers all components in the optimisation problem, from PDEs to objective functions, as mathematical functions that can be arbitrarily composed, and whose derivative can be automatically calculated through the procedure presented above. In Stride, each of these functions is a \texttt{stride.Operator} object, where their inputs and outputs are \texttt{stride.Variable} objects (Listing \ref{lst:1}).

\begin{lstlisting}[language=Python3, float=h, label={lst:1}, caption={Example calculation of the gradient of a chain of functions using Stride. Note the use of the \texttt{await} syntax that is needed for compatibility with the Mosaic parallelisation library.}]
from stride import Variable

x = Variable(name="x", 
             needs_grad=True)
z = await h(x)
y = await g(z)
w = await f(y)

await w.adjoint()
# The gradient is now in "x.grad"
\end{lstlisting}

When each \texttt{stride.Operator} is called, it is immediately applied on its inputs to generate some outputs. At the same time, these outputs keep a record of the chain of calls that have led to them within a directed acyclic graph. When \texttt{w.adjoint()} is called, this graph is traversed from the root \texttt{w} to the leaf \texttt{x}, calculating the gradient in the process. Only the leaves for which the flag \texttt{needs\_grad} is set to \texttt{True} will have their gradient computed, which will be stored in the internal buffer of the variable \texttt{x.grad}.

Now, we proceed to apply these general abstractions to find the gradient of a more practical optimisation problem. Consider the PDE-constrained optimisation problem,

\begin{equation}
\begin{split}
    \mathbf{m}^* = \argmin_{\mathbf{m}} & J(\mathbf{u}, \mathbf{m}) = 
    \argmin_{\mathbf{m}} \left\langle \hat{J}(\mathbf{u}, \mathbf{m}), 1 \right\rangle \\
    &s.t.\; \mathbf{L}(\mathbf{u},\mathbf{m}) = \mathbf{0}
\end{split}
\end{equation}

\noindent
given some scalar objective function or loss function $J(\mathbf{u}, \mathbf{m})$ and some PDE $\mathbf{L}(\mathbf{u},\mathbf{m}) = \mathbf{0}$, for some vector of state variables $\mathbf{u}$ and a vector of design variables $\mathbf{m}$. Considering $\mathbf{L}(\mathbf{u},\mathbf{m})$ to be an adequate, continuously differentiable function in some neighbourhood of $\mathbf{m}$, we can apply the implicit function theorem. Then $\mathbf{L}(\mathbf{u},\mathbf{m}) = \mathbf{0}$ has a unique continuously differentiable solution $\mathbf{u}(\mathbf{m})$ and its derivative is given by the solution of,

\begin{equation}
\begin{split}
    \nabla_\mathbf{u}\mathbf{L}(\mathbf{u}(\mathbf{m}), \mathbf{m}) \nabla_\mathbf{m}\mathbf{u}(\mathbf{m}) \delta\mathbf{m} +
    \nabla_\mathbf{m}\mathbf{L}(\mathbf{u}(\mathbf{m}), \mathbf{m}) \delta\mathbf{m} = \mathbf{0} \\
    \nabla_\mathbf{m}\mathbf{u}(\mathbf{m})\delta\mathbf{m} = - \nabla_\mathbf{u}\mathbf{L}^{-1}(\mathbf{u}(\mathbf{m}), \mathbf{m})
    \nabla_\mathbf{m}\mathbf{L}(\mathbf{u}(\mathbf{m}), \mathbf{m}) \delta\mathbf{m}
\end{split}
\label{eq:implicit}
\end{equation}

We can then define a reduced objective $F(\mathbf{m}) = J(\mathbf{u}(\mathbf{m}), \mathbf{m}) = \left\langle \hat{J}(\mathbf{u}(\mathbf{m}), \mathbf{m}), 1 \right\rangle$, and we can take its derivative with respect to $\mathbf{m}$ by using the previously introduced procedure,

\begin{equation}
\begin{split}
    \nabla_\mathbf{m} & F(\mathbf{m})(\delta \mathbf{m}) = 
    \left\langle \nabla_\mathbf{u}\hat{J}(\mathbf{u}(\mathbf{m}), \mathbf{m}), \nabla_\mathbf{m}\mathbf{u}(\mathbf{m})\delta\mathbf{m} \right\rangle \\
    &+ \left\langle \nabla_\mathbf{m}\hat{J}(\mathbf{u}(\mathbf{m}), \mathbf{m}), \delta \mathbf{m} \right\rangle \\
    &= \left\langle \nabla_\mathbf{m}^*\mathbf{u}(\mathbf{m}) \nabla_\mathbf{u}\hat{J}(\mathbf{u}(\mathbf{m}), \mathbf{m}), \delta\mathbf{m} \right\rangle \\
    &+ \left\langle \nabla_\mathbf{m}\hat{J}(\mathbf{u}(\mathbf{m}), \mathbf{m}), \delta \mathbf{m} \right\rangle
\end{split}
\label{eq:diff}
\end{equation}

Substituting expression \ref{eq:implicit} into expression \ref{eq:diff} we obtain,

\begin{equation}
\begin{split}
    \nabla_\mathbf{m} & F(\mathbf{m})(\delta \mathbf{m}) = 
    \left\langle \nabla_\mathbf{m}^*\mathbf{u}(\mathbf{m}) \nabla_\mathbf{u}\hat{J}(\mathbf{u}(\mathbf{m}), \mathbf{m}), \delta\mathbf{m} \right\rangle \\
    &+ \left\langle \nabla_\mathbf{m}\hat{J}(\mathbf{u}(\mathbf{m}), \mathbf{m}), \delta \mathbf{m} \right\rangle \\
    &= - \left\langle \nabla_\mathbf{m}\mathbf{L}^*(\mathbf{u}(\mathbf{m}), \mathbf{m})
    \nabla_\mathbf{u}\mathbf{L}^{-*}(\mathbf{u}(\mathbf{m}), \mathbf{m}) \right. \\
    & \left. \nabla_\mathbf{u}\hat{J}(\mathbf{u}(\mathbf{m}), \mathbf{m}), \delta\mathbf{m} \right\rangle \\
    &+ \left\langle \nabla_\mathbf{m}\hat{J}(\mathbf{u}(\mathbf{m}), \mathbf{m}), \delta \mathbf{m} \right\rangle \\
    &= \left\langle \nabla_\mathbf{m}\mathbf{L}^*(\mathbf{u}(\mathbf{m}), \mathbf{m}) \mathbf{w}(\mathbf{m}), \delta\mathbf{m} \right\rangle \\
    &+ \left\langle \nabla_\mathbf{m}\hat{J}(\mathbf{u}(\mathbf{m}), \mathbf{m}), \delta \mathbf{m} \right\rangle
\end{split}
\end{equation}

\noindent
where $\mathbf{w}(\mathbf{m})$ is the solution of the adjoint PDE,

\begin{equation}
    \mathbf{w}(\mathbf{m}) = 
    - \nabla_\mathbf{u}\mathbf{L}^{-*} (\mathbf{u}(\mathbf{m}), \mathbf{m})
    \nabla_\mathbf{u}\hat{J}(\mathbf{u}(\mathbf{m}), \mathbf{m})
\end{equation}

In this optimisation problem, both $\mathbf{L}(\mathbf{u}, \mathbf{m})$ and $J(\mathbf{u}, \mathbf{m})$ would be \texttt{stride.Operator} objects. Adding new functions to Stride requires defining a new \texttt{stride.Operator} subclass that implements two methods, \texttt{forward} and \texttt{adjoint} (Listing \ref{lst:2}).

\begin{lstlisting}[language=Python3, float=h, label={lst:2}, caption={Example of gradient calculation for a PDE-constrained optimisation problem like the one solved in FWI.}]
from stride import Operator, Variable

class L(Operator):
    def forward(self, m):
        # Compute wave equation solution
        return u
        
    def adjoint(self, grad_u, m):
        # Calculate derivative wrt to m
        # applying adjoint on grad_u
        return grad_m
        
class J(Operator):
    def forward(self, u, m):
        # Calculate loss value
        return loss
        
    def adjoint(self, grad_loss, u, m):
        # Calculate the derivative wrt u
        # Calculate the derivative wrt m
        return grad_u, grad_m
        
# Create the design parameters
m = Variable(name="m")
m.needs_grad = True

# Instantiate the operators
l = L()
j = J()

# Apply to calculate gradient
u = await l(m)
loss = await j(u, m)

await loss.adjoint()
# The gradient is now in "m.grad"
\end{lstlisting}

\begin{lstlisting}[language=Python3, float=b, label={lst:3}, caption={Once a gradient has been calculated, a step in the optimisation algorithm can be taken by using a \texttt{stride.Optimiser}.}]
from stride import GradientDescent

optimiser = GradientDescent(m, step_size=1.)
await optimiser.step()
\end{lstlisting}

The abstractions presented allow us to intuitively pose optimisation problems and calculate derivatives of an objective function with respect to the parameters of interest. However, in order to solve the problem, we have to apply this derivative to update our guess of the parameters and repeat the procedure iteratively until we are satisfied with the final result.

Stride provides local optimisers of type \texttt{stride.LocalOp-\\timiser} that determine how parameters should be updated given an available derivative. For our previous example, we can follow the procedure in Listing \ref{lst:3} to apply a step of gradient descent in the direction of our calculated derivative. Writing new, user-defined optimisers only requires the creation of a \texttt{stride.LocalOptimiser} subclass that takes the \texttt{stride.Variable} being optimised when the class is instantiated and that defines the method \texttt{step()}, which executes a single step in the optimisation process.

\begin{lstlisting}[language=Python3, float=t, label={lst:4}, caption={Running through multiple iterations in the optimisation can be easily structured using the \texttt{stride.OptimisationLoop}.}]
from stride import OptimisationLoop

opt_loop = OptimisationLoop()

for block in opt_loop.blocks(num_blocks):
    for iteration in \
            block.iterations(num_iters):
        m.clear_grad()
        
        u = await l(m)
        loss = await j(u, m)
        await loss.adjoint()
        
        await optimiser.step()
\end{lstlisting}

In order to iterate through the optimisation procedure, we could use a standard Python \texttt{for} loop. However, we also provide in Stride a \texttt{stride.OptimisationLoop} to use in these cases, which will help structure and keep track of the optimisation process.

Iterations in Stride are grouped together in blocks, with the \texttt{stride.OptimisationLoop} containing multiple blocks and each block containing multiple iterations. Partitioning the inversion in this way allows us to divide the optimisation more easily into logical units that share some characteristics. For instance, in FWI it is common to gradually introduce frequency information into the inversion to better condition the optimisation. In this case, it would make sense to assign one block to each frequency band, and run that band for some desired number of iterations. Listing \ref{lst:4} adds a \texttt{stride.OptimisationLoop} around our previous example.

\subsection{Problem definition}

In addition to providing abstractions for solving optimisation problems, Stride introduces a series of utilities for users to specify the characteristics of the problem being solved, such as the physical properties of the medium or the sequence in which transducers are used.

In Stride, the problem is first defined over a spatiotemporal grid, which determines the spatial and temporal bounds of the problem and their discretisation (Listing \ref{lst:5}). Currently, we support discretisations over rectangular grids, but other types of meshes could be introduced in the future. On this spatiotemporal mesh, we define a series of grid-aware data containers, which include scalar and vector fields, and time traces. These data containers are subclasses of \texttt{stride.Variable}.

\begin{lstlisting}[language=Python3, float=h, label={lst:5}, caption={Example spatiotemporal grid.}]
from stride import Space, Time, Grid

space = Space(shape, spacing)
time = Time(start, step, num)

grid = Grid(space, time)
\end{lstlisting}

Based on this, we can define a medium, a \texttt{stride.Medium} object, a collection of fields that determine the physical properties in the region of interest. For instance, the medium could be defined by two \texttt{stride.ScalarField} objects containing the spatial distribution of longitudinal speed of sound and density, as in Listing \ref{lst:6}.

\begin{lstlisting}[language=Python3, float=h, label={lst:6}, caption={Example \texttt{stride.Medium} containing the spatial distribution of longitudinal speed of sound and density.}]
from stride import Medium, ScalarField

medium = Medium(grid=grid)
medium.add(ScalarField(name="vp",
                       grid=grid))
medium.add(ScalarField(name="rho", 
                       grid=grid))

medium.vp.fill(1500.)
medium.rho.fill(1000.)
\end{lstlisting}

Next, we can define the transducers, the computational representation of the physical devices that are used to emit and receive sound, characterised by aspects such as their geometry and impulse response. These transducers are then located within the spatial grid by defining a series of locations in a \texttt{stride.Geometry}. In Listing \ref{lst:7} we instantiate some \texttt{stride.Transducer} objects and then add them to a corresponding \texttt{stride.Geometry}.

\begin{lstlisting}[language=Python3, float=h, label={lst:7}, caption={Example geometry with its associated transducers.}]
from stride import PointTransducer, \
                   Transducers, Geometry

transducers = Transducers(grid=grid)
trans_0 = PointTransducer(0, grid=grid)
trans_1 = PointTransducer(1, grid=grid)

transducers.add(trans_0)
transducers.add(trans_1)

geometry = Geometry(transducers=transducers,
                    grid=grid)
geometry.add(0, 
             transducer=trans_0, 
             coordinates=[...])
geometry.add(1, 
             transducer=trans_1, 
             coordinates=[...])
\end{lstlisting}

Finally, we can specify an acquisition sequence within a \texttt{stride.Acquisitions} object (Listing \ref{lst:8}). The acquisition sequence is composed of shots (\texttt{stride.Shot} objects), where each shot determines which transducers at which locations act as sources and/or receivers at any given time during the acquisition process. The shots also contain information about the wavelets used to excite the sources and the data observed by the corresponding receivers if this information is available.

\begin{lstlisting}[language=Python3, float=h, label={lst:8}, caption={Example acquisition containing only one shot.}]
from stride import Shot, Acquisitions

loc_0 = geometry.get(0)
loc_1 = geometry.get(1)

acquisitions = Acquisitions(geometry=geometry,
                            grid=grid)
shot = Shot(0, 
            sources=[loc_0], 
            receivers=[loc_0, loc_1],
            geometry=geometry,
            grid=grid)
acquisitions.add(shot)
\end{lstlisting}

All components of the problem definition can be stored in a \texttt{stride.Problem} object, which structures them in a single, common entity.

\subsection{Physical modelling}

Physical modelling is defined in Stride through \texttt{stride.\newline Operator} objects that represent specific implementations of a numerical solver applied to a PDE. Stride does not prescribe a specific solver or numerical method, and different codes and implementations can be integrated with it as long as they conform to the \texttt{stride.Operator} interface.

By default, Stride integrates with the Devito library, a domain-specific language that generates highly optimised finite-difference code from high-level symbolic differential equations \cite{Louboutin2019DevitoExploration, Luporini2020ArchitectureComputation}. Using Devito, we provide an out-of-the-box implementation of the second-order isotropic acoustic wave equation, for which Devito automatically generates code that can be readily executed in parallel on CPUs using Open Multi-Processing (OpenMP), and on GPUs using both OpenMP and OpenACC.

Acoustic modelling in Stride is governed by the equation,

\begin{equation}
    \frac{1}{v_p^2} \frac{\partial^2 p}{\partial t^2} = \rho \nabla \cdot (\frac{1}{\rho} \nabla p) + \eta \frac{\partial}{\partial t} (-\nabla^2)^{y/2} p
\end{equation}

\noindent
where $p(t, \mathbf{x})$ is the pressure, $v_p(\mathbf{x})$ is the longitudinal speed of sound, $\rho(\mathbf{x})$ is the mass density, $\eta = -2 \alpha_0 v_p^{y-1}$, and $\alpha_0(\mathbf{x})$ is the absorption coefficient. The implementation of the acoustic wave equation is fourth-order accurate in time and tenth-order accurate in space. This results in a stability region with Courant-Friedrichs-Lewy (CFL) constant of 0.81 in 2D and 0.66 in 3D \cite{Amundsen2017TimeEquations}, as well as the requirement of a minimum of 3 points per wavelength (PPW) to minimise numerical dispersion. Our solver includes options for both constant and variable density and attenuation. Attenuation follows a power law, with frequency dependence controlled by the parameter $y$ in the equation, which can take values 0 and 2. In these cases the implemented derivative is not fractional.

\begin{lstlisting}[language=Python3, float=t, label={lst:9}, caption={Basic usage of Mosaic to create remote objects, call their methods and access their attributes.}]
from mosaic import tessera

@tessera
class Remote:
    def __init__(self):
        self.value = 0
    def add(self, value):
        self.value += value
        return self.value

# Create a remote instance        
remote_obj = Remote.remote()

# Check the current value of the attribute
print(await remote_obj.value)

# Add a new value
task = remote_obj.add(5)

# This will return immediately and 
# we can do other work in the meantime

# When ready, we can wait for the
# remote method call to finish
await task

# The return value of the method call 
# is stored in the remote worker, and 
# to access them we will need to do 
# this explicitly
print(await task.result())

# Check the new value of the attribute
print(await remote_obj.value)
\end{lstlisting}

\begin{lstlisting}[language=Python3, float=t, label={lst:10}, caption={Expressing parallelism and dependencies in Mosaic.}]
# Create a remote instances    
remote_obj_0 = Remote.remote()
remote_obj_1 = Remote.remote()

# These could be executed in parallel
task_0 = remote_obj_0.add(5)
task_1 = remote_obj_1.add(1)

# and have to be awaited separately
await task_0
await task_1

# An explicit dependence will make
# them execute in series
task_0 = remote_obj_0.add(1)
task_1 = remote_obj_1.add(task_0)

# and only the latter needs to be awaited
print(await task_1.result())  # will print 7

# Dependencies can also be introduced as
task_0 = remote_obj_0.add(1)
task_1 = remote_obj_1.add(task_0.done, 2)

await task_1
\end{lstlisting}

In terms of boundary conditions, Stride includes options for a sponge absorbing boundary \cite{Yao2018AnSimulation} or a perfectly matched layer \cite{Gao2015UnsplitEquations}. In all cases, sources and receivers can be defined in locations off the grid, with both bi-/tri-linear interpolation and high-order sinc interpolation \cite{Hicks2002ArbitraryFunctions}. It is important to note that current, out-of-the-box implementations of the adjoints of our PDE solvers consider domains to be unbounded, as these represent the most common scenario in ultrasound imaging. However, alternative boundary conditions can be readily accounted for through user-level extensions of the PDE operators.

Although physical modelling in Stride is currently focused on finite-difference methods, future releases could include integration with pseudospectral-element DSLs such as Dedalus \cite{Burns2020Dedalus:Methods} or finite-element DSLs like FEniCS/Firedrake \cite{Logg2012AutomatedEngineering,Rathgeber2016Firedrake:Abstractions}.

\subsection{Parallelism}

In practice, derivatives of the optimisation problem are not calculated one data point at a time, but in batches, and the result is averaged to obtain an estimate of the gradient for that iteration. Because, in most cases, each of these data points is fully independent, this can be exploited so that they are calculated in parallel. For some simple problems, this can be done within a single workstation. However, in most practical problems, compute and memory demands require that these computations are mapped across different interconnected sets of hardware, such as multi-GPU systems and CPU clusters, running locally, remotely, or on the cloud.

The most important limiting factor when scaling real-life FWI workloads in parallel environments is memory allocation, management, and communication, with potentially hundreds of gigabytes being stored and transferred during the optimisation process. Therefore, a parallelisation framework is required that offers fine-grained control of the computational workload allocation and memory management for code developers, while also providing the end user with a high level of abstraction that integrates tightly with the optimisation constructs provided by Stride. We have developed Mosaic to facilitate the expression of parallelism in Stride in an intuitive manner.

Mosaic is an actor-based parallelisation library based on asynchronous, zero-copy message passing through ZeroMQ sockets \cite{ZeroMQDevelopmentTeamZeroMQ:Library}. Actors in Mosaic are called tessera, and can be generated by decorating any Python class using \texttt{@mosaic.tessera}. When instantiating a class that has been decorated, Mosaic will start a remote instance of that class in one of the workers. At this point, remote method calls to that tessera can be executed and the attributes of that remote object can be accessed. An example of how Mosaic is used can be found in Listing \ref{lst:9}.

In Mosaic, subsequent method calls to a remote object are guaranteed to be executed in order, but calls to different remote objects are not. However, if there are explicit dependencies between two or more remote method calls, Mosaic will ensure that these are executed in the right order (Listing \ref{lst:10}).

\begin{figure}[t]
    \centering
    \includegraphics[width=8.2cm]{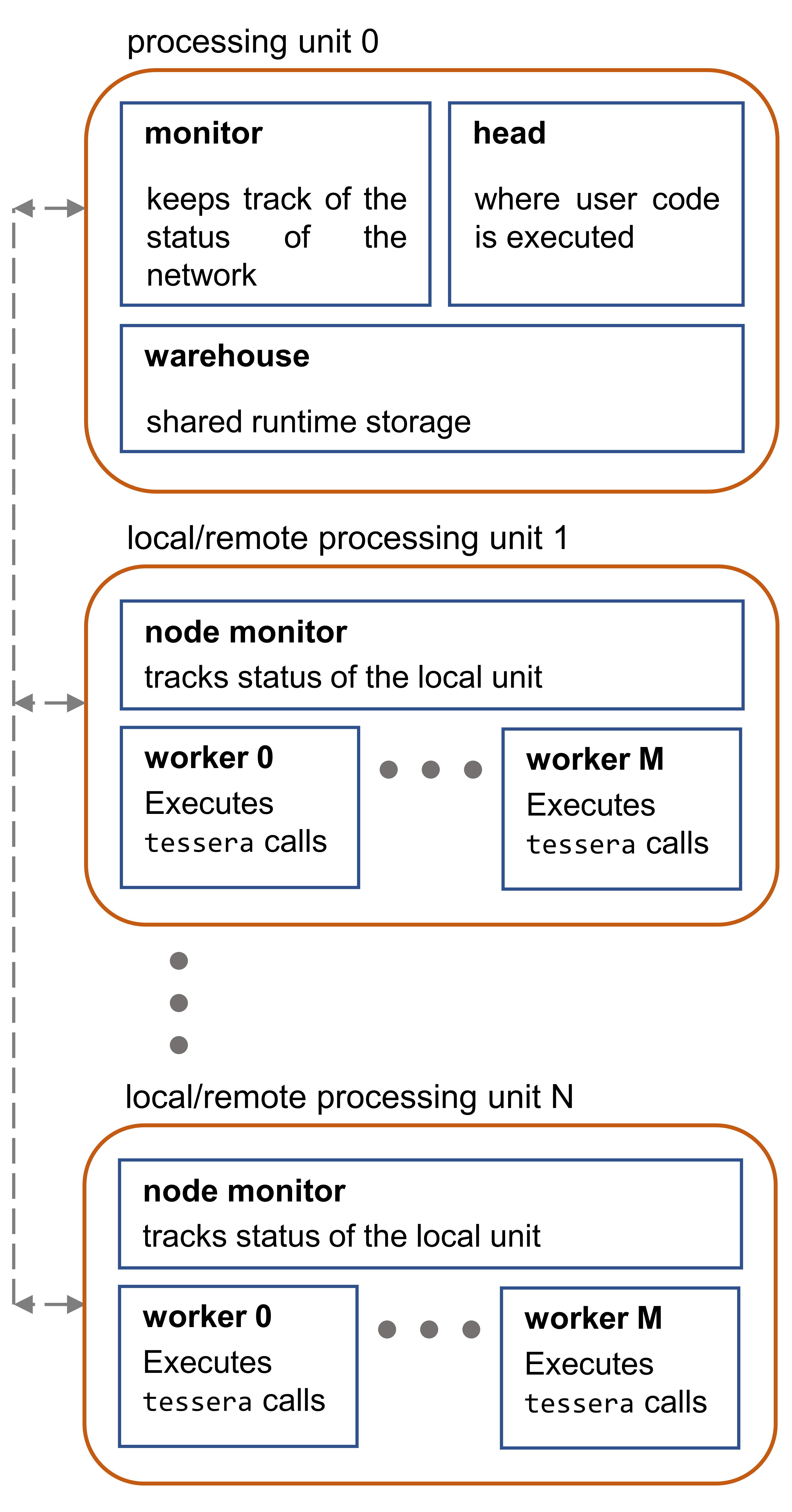}
    \caption{Schematic representation of the Mosaic runtime. The runtime is divided into several logical processing units, which could represent, for instance, processes in a local multi-processing environment or different machines in a multi-node cluster. In the first processing unit, the user code is executed in the \textit{head}, while the \textit{monitor} tracks the status of the runtime and the \textit{warehouse} acts as a central storage unit. In the remaining processing units, a \textit{node monitor} is allocated to track the status of that local unit and communicate this to the global \textit{monitor}, and one or more \textit{workers} are also created to execute \texttt{tessera} calls. All endpoints in the Mosaic runtime are interconnected to each other.}
    \label{fig:mosaic}
\end{figure}

The structure of the Mosaic runtime, which can be seen in Fig. \ref{fig:mosaic}, is composed by a series of processing units, which could be located in a single, local workstation or distributed across a remote network. The first of such units contains a \textit{monitor} process, a \textit{warehouse} process, and a \textit{head} process. The \textit{monitor} process collects information about the Mosaic network, including occupation rate, resource use and connection state. The \textit{warehouse} process acts as a centralised key-value storage location that is accessible from across the whole Mosaic network. The \textit{head} process is the place where the main user code is executed. In each of the remaining processing units, a \textit{node monitor} and one or more \textit{workers} are allocated. The \textit{node monitor} keeps track of the runtime status of its local processing unit and oversees the life cycle of each of the \textit{workers} in its unit. Finally, the \textit{workers} act as containers for tessera actors, whose methods can be executed remotely. All processing units in the Mosaic network are directly interconnected to each other, creating a decentralised communication mesh.

Mosaic can be run in interactive mode in a Jupyter notebook, or from a terminal window using the \texttt{mrun} command. The Mosaic runtime can be used without any code changes in a local multi-processing environment or a multi-node cluster. Therefore, Mosaic gives us the flexibility to parallelise work across multiple CPUs within a single compute node, as well as across multiple interconnected nodes, with the distribution topology related to the specific application at hand. Additionally, our Devito solvers can parallelise the execution of the wave equation across multiple CPU cores by using thread-level parallelism.

\subsection{File input and output}

\begin{figure*}[t]
    \centering
    \includegraphics[width=17.5cm]{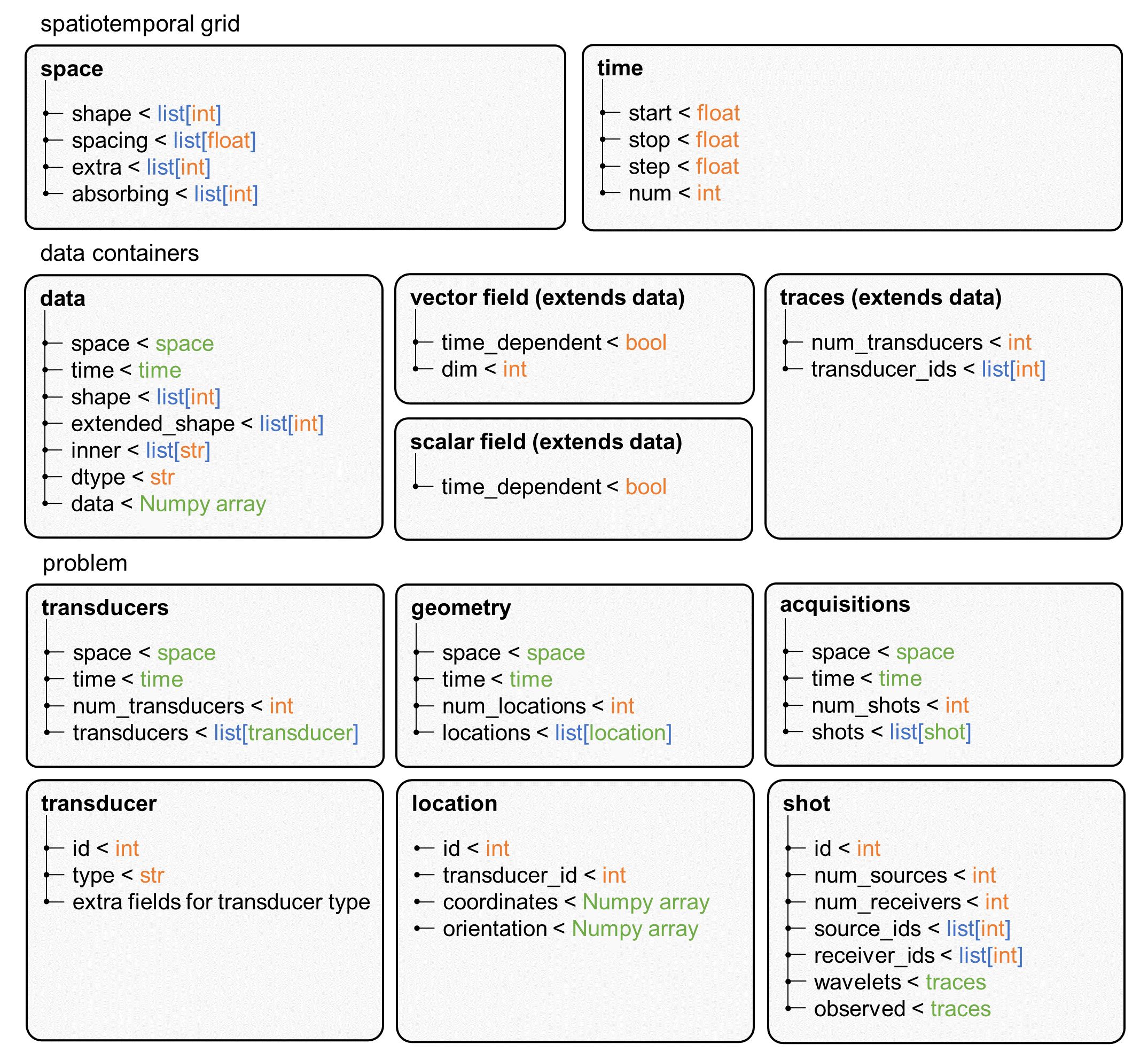}
    \caption{Specification of the Stride file format. The definition of the spatiotemporal grid is the basis upon which different types of data containers and the various components of the problem are then specified.}
    \label{fig:file}
\end{figure*}

\begin{figure*}[t]
    \centering
    \includegraphics[width=17.5cm]{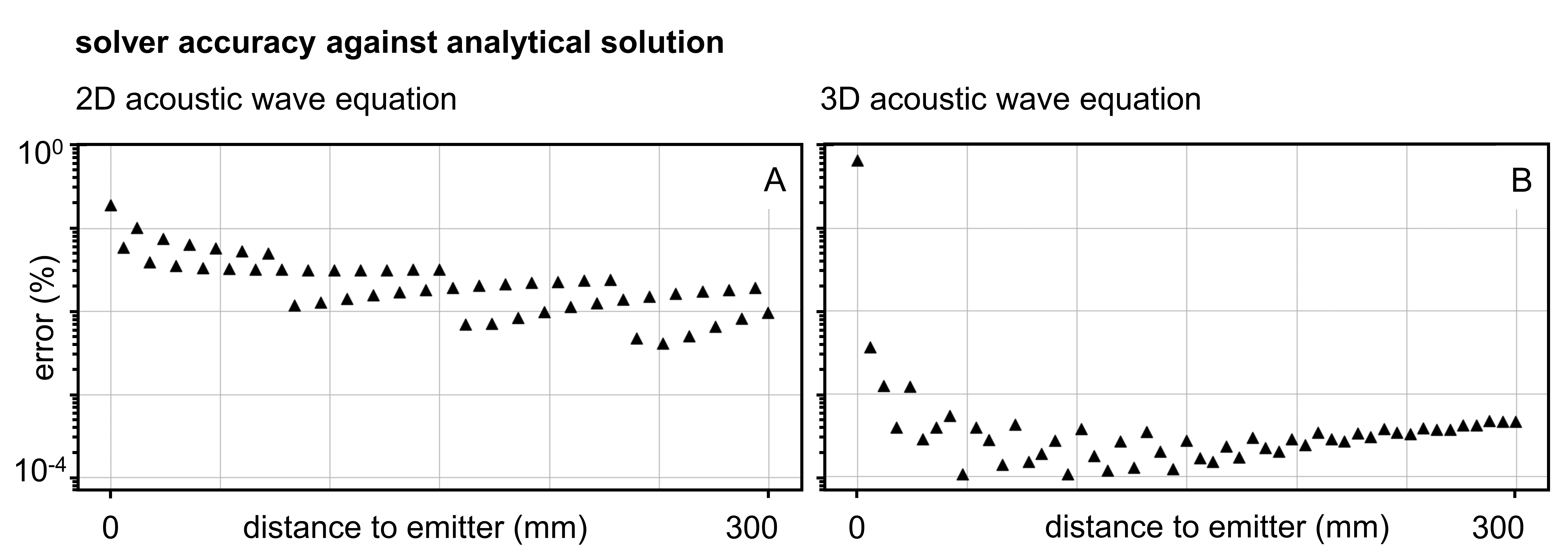}
    \caption{Accuracy of the acoustic wave equation solver against analytical solution. The numerical solution of the acoustic wave equation calculated by Stride is compared to the analytical solution for a medium with homogeneous speed of sound. The comparison is performed in 2D (A) and 3D (B), at a distance to the emitter ranging from 0 to 300 mm. Error is calculated as the normalised root-mean-square error with respect to the analytical solution.}
    \label{fig:analytic}
\end{figure*}

\begin{figure*}[htp]
    \centering
    \includegraphics[width=17.5cm]{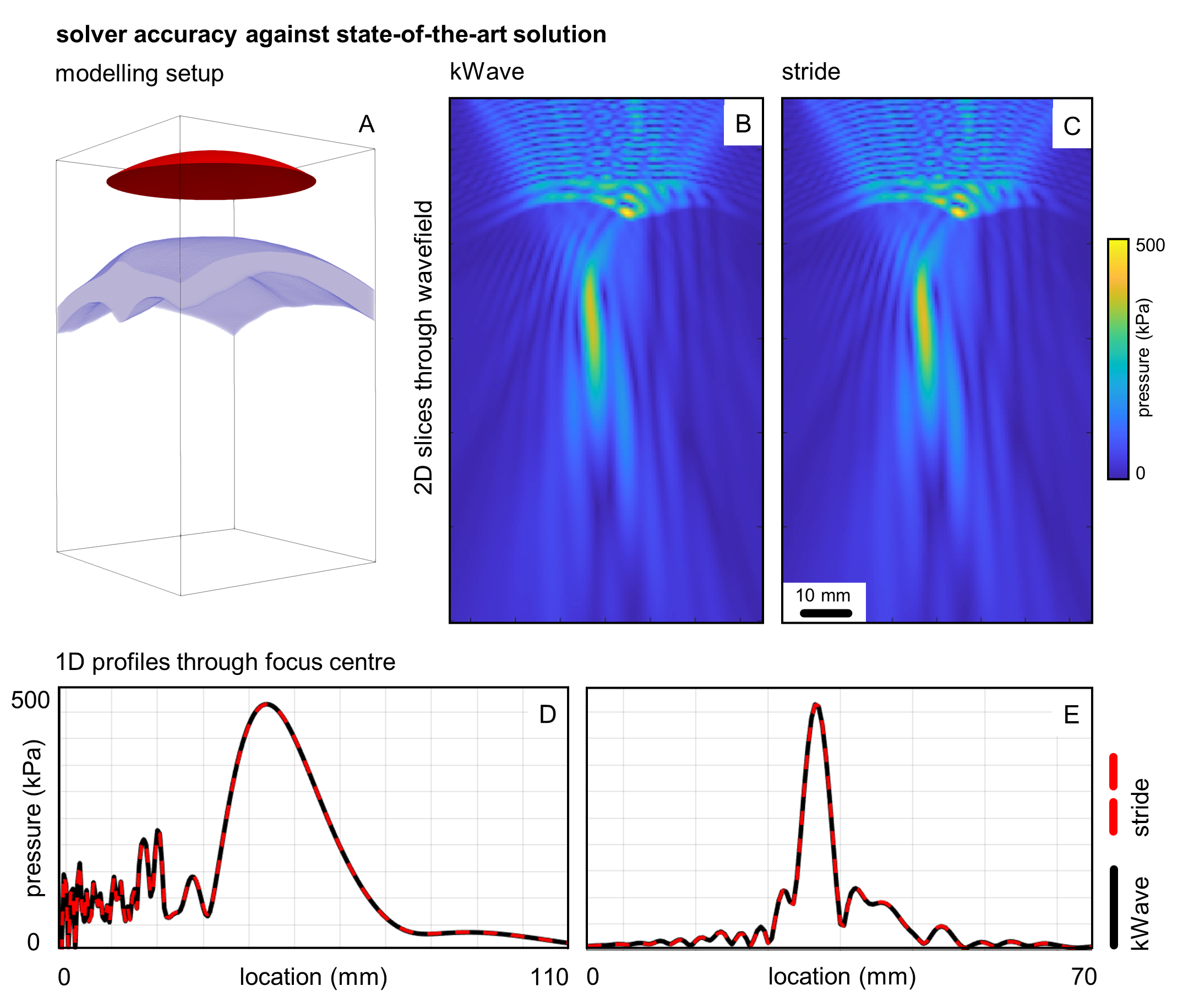}
    \caption{Accuracy of the acoustic wave equation solver against state-of-the-art solver. The 3D numerical model (A) contains a human skull section (blue) and a bowl ultrasound transducer (red). We compare a 2D slice through the resulting steady-state wavefield for the state-of-the-art solver kWave (B) and for Stride (C). Additionally, we compare two 1D profiles through the centre of the transducer focus (D-E).}
    \label{fig:kWave}
\end{figure*}

As the popularity of ultrasound tomography increases, the number and size of datasets are also growing, but no standard format exists for their exchange. This slows algorithm development and limits research reproducibility. In order to address this, we have introduced with Stride a standardised file specification and a set of tools to interact with it.

In the setup of ultrasound tomography workflows, there are usually a number of intermediate files that are generated describing aspects such as medium properties, transducer impulse responses or data recorded during laboratory experiments. In Stride, we use the Hierarchical Data Format (HDF5)
\cite{TheHDFGroupHierarchical5} for saving and loading these datasets and provide a series of tools to conveniently interact with them. Fig. \ref{fig:file} shows the basic file specification proposed in Stride for the different components of a standard tomographic workflow.

\section{Results}
\label{sec:results}

\subsection{Modelling accuracy}
\label{sec:accuracy}

We have validated the accuracy of the acoustic solver by comparing it against an analytical solution of the wave equation for a homogeneous medium \cite{Morse1953MethodsPhysics}. The comparison was performed, in 2D and 3D, by transmitting a three-cycle tone burst centred at 500 kHz into a medium with constant speed of sound of 1500 m/s, constant density, and no attenuation. The employed grid was sampled at 0.250 mm in space (minimum of 8 PPW) and \ilu[0.060]{\micro\second} in time (maximum CFL constant of 0.36). The resulting acoustic wave was then recorded at 51 equispaced points, starting at the transmission location and increasing in distance up to a maximum separation of 300 mm.

Results for the comparison are shown in Fig. \ref{fig:analytic}, both for the 2D (Fig. \ref{fig:analytic}-A) and the 3D cases (Fig. \ref{fig:analytic}-B), where errors with respect to the analytical solution were calculated using the normalised root-mean-square error. We can see how the Stride numerical solutions closely match the analytical ones, remaining accurate at a significant distance from the transmission site.

We have performed a further validation of the Stride acoustic solvers on a more complex medium with inhomogeneous speed of sound, density, and attenuation of order zero, for which an analytical solution is not available, by comparing it against kWave \cite{Treeby2010K-Wave:Fields}, a state-of-the-art ultrasound modelling library written in MATLAB and based on pseudospectral element methods. The comparison was performed using a human skull section, seen in Fig. \ref{fig:kWave}-A, sampled at 0.125 mm (minimum of 24 PPW), and illuminated by a bowl ultrasound transducer with a 64 mm radius of curvature and a 64 mm aperture diameter. The transducer surface was discretised using 20,000 point sources, evenly distributed using Fibonacci spirals \cite{Vogel1979AHead}. This example forms part of a transcranial ultrasound simulation benchmarking and intercomparison exercise organised by the ITRUSST (International Transcranial Ultrasonic Stimulation Safety and Standards) planning group \cite{Aubry2022BenchmarkModels}. The transducer was excited by a continuous sinusoidal wave at 500 kHz and the simulation was run with a step size of \ilu[0.016]{\micro\second} (maximum CFL constant of 0.36) until steady state was reached. The magnitude of the pressure field at the excitation frequency was then extracted after Fourier transform. Fig. \ref{fig:kWave}-B and C show a 2D slice through the resulting 3D wavefield, from which we can observe the good agreement between both solutions. A similar conclusion can be extracted from the 1D profiles, seen in Fig. \ref{fig:kWave}-D and E. The agreement between both solvers is quantitatively confirmed by a relative error of 1.64\%, calculated over the entire 3D volume. Existing differences between the results of both solvers are likely due to the use of different numerical methods to solve the wave equation, as well as differences in source injection routines and boundary condition implementation. It is important to note that implementation differences cannot be fully eliminated, even in the limit where both numerical methods converge, due to the fact that Stride and kWave are solving fundamentally different equations in order to model acoustic wave propagation: Stride solves the second-order linear acoustic wave equation, whereas kWave solves three coupled equations that are equivalent to a generalized Westervelt equation.

\subsection{Imaging in 2D}

\begin{lstlisting}[language=Python3, float=b, label={lst:11}, caption={To image the spatial distribution of speed of sound, we create a \texttt{stride.ScalarField(..., needs\_grad=True)} and set the starting distribution to be 1500 m/s everywhere. We also create a \texttt{stride.GradientDescent} optimiser to update the variable at every iteration.}]
# Prepare starting model
vp = ScalarField.parameter(name="vp", 
                           grid=grid, 
                           needs_grad=True)
vp.fill(1500.)
medium.add(vp)

# Prepare optimiser
optimiser = GradientDescent(vp, 
                    step_size=step_size)
\end{lstlisting}

\begin{lstlisting}[language=Python3, float=b, label={lst:12}, caption={We create the necessary operators for the reconstruction. The keyword argument \texttt{len=num\_workers} controls the amount of copies of the operators to be instantiated by Mosaic in each remote worker.}]
# Prepare operators
pde = IsoAcousticDevito.remote(grid=grid, 
                            len=num_workers)
loss = L2DistanceLoss.remote(
                            len=num_workers)
p_wavelets = ProcessWavelets.remote(
                            len=num_workers)
p_traces = ProcessTraces.remote(
                            len=num_workers)
\end{lstlisting}

\begin{figure}
    \centering
    \includegraphics[width=8.5cm]{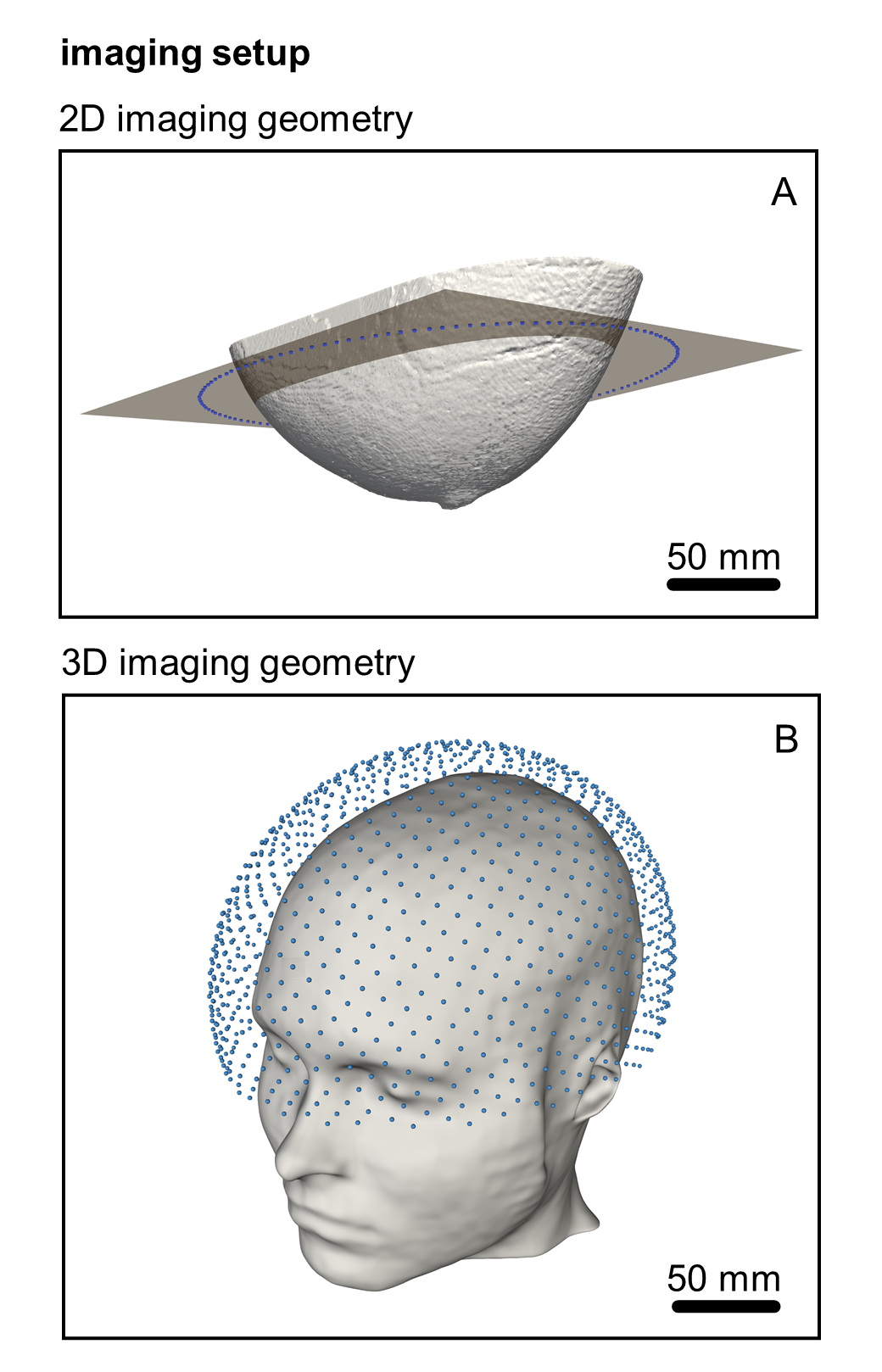}
    \caption{Setup used in the numerical experiments. For the 2D experiment (A), a slice is taken across a numerical 3D model of the breast and 128 point transducer, which can be seen as blue dots, are distributed around it. For the 3D experiment (B), a numerical head model is imaged by surrounding it with 1024 transducers (also visible as blue dots). The scale shown at the bottom of the numerical models applies equally in all spatial directions.}
    \label{fig:geometry}
\end{figure}

\begin{figure}[t]
    \centering
    \includegraphics[width=8.5cm]{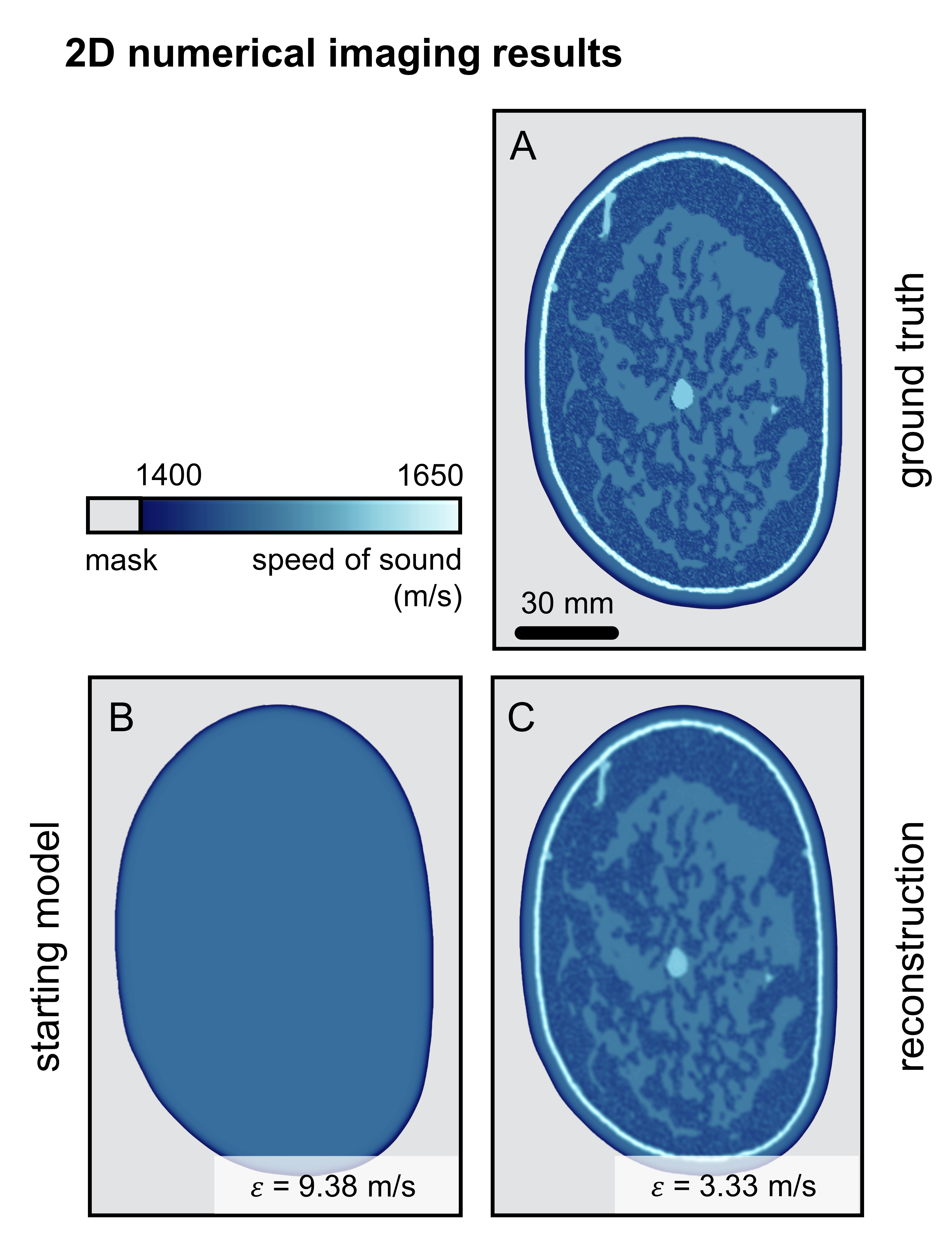}
    \caption{Stride reconstruction in 2D. A 2D acoustic breast model (A) is imaged starting from a homogeneous distribution of speed of sound (B). Stride manages to accurately reconstruct the target model (C). The mean of the absolute value of the difference between the ground-truth model and the inversion is displayed here as $\varepsilon$.}
    \label{fig:breast2D}
\end{figure}

For our first imaging experiment, we extract a 2D slice from a numerical breast model as seen in Fig. \ref{fig:geometry}-A. The resulting 2D model can be seen in Fig. \ref{fig:breast2D}-A. The model has been obtained from an open database \cite{Lou2017GenerationImaging}, and has been adapted by populating it with acoustic tissue properties and by adding a tumour. From here onwards, all examples were run with constant density and no attenuation. The model, sampled with a spacing of 0.500 mm (minimum of 3.73 PPW), has a size of 456\texttimes 485. The model is surrounded by 128 point transducers, seen as blue dots in Fig. \ref{fig:geometry}-A, all of which act as sources and receivers. Imaging is performed using a three-cycle tone burst centred at 500 kHz, and is carried out over \ilu[200]{\micro\second} in steps of \ilu[0.080]{\micro\second} (maximum CFL constant of 0.26). Both temporal and spatial sampling are kept constant during modelling and inversion, for this and all subsequent examples. However, this is not required and users could exploit different dispersion and stability conditions by changing the discretisation across different imaging blocks.

To make use of the gradient-calculation capabilities of Stride, we instantiate our speed-of-sound field with \texttt{needs\_grad=True}, and set the starting model to a constant sound speed of 1500 m/s (Fig. \ref{fig:breast2D}-B). We also instantiate a gradient descent optimiser to update our variable (Listing \ref{lst:11}).

We can see in Listing \ref{lst:11} how the \texttt{stride.ScalarField} has been instantiated by calling \texttt{parameter()}. Using this method will ensure that, as the field is sent across the Mosaic network, a reference to the original object will always be maintained. This will allow us to calculate the gradient in different workers and then propagate the results back to the local runtime.

Then, we can instantiate our operators remotely, creating one copy for each available worker (Listing \ref{lst:12}). In this case, we use an operator for the PDE and another one for the objective function, and we also create pre-processing operators for our source wavelets and our output time traces.

We perform the inversion by gradually introducing frequencies, starting at 300 kHz and going up to 600 kHz. We do this by running the optimisation loop in blocks, with each block using a different frequency band. At each block, we complete 8 iterations, randomly selecting 16 shots without replacement in each of them. That is, each shot is used once at every frequency band. We run the function in Listing \ref{lst:13} for every iteration of the reconstruction loop in Listing \ref{lst:14}. We run this inversion on a local multi-processing environment, within a Jupyter notebook, by simply adding the command \texttt{mosaic.interactive("on")} at the beginning of our notebook. This workstation is equipped with 64 GB of memory and 6 physical cores (Intel i7-8700K, 6 cores, 3.70 GHz). The acoustic Devito PDE was compiled using the GNU \texttt{gcc} compiler version 7.5, and was executed on the Jupyter notebook using 3 Mosaic workers and OpenMP thread-level parallelism with 2 threads for each worker. Each of the Mosaic workers calculates the gradient for a single shot at a time, which entails one forward propagation and one adjoint propagation of the acoustic solver, before combining the gradients for all shots at each iteration. With this configuration, each shot gradient calculation took a total of 2.99 \textpm\ 0.30 s.

Once the optimisation loop runs through all frequency bands, a final reconstruction is obtained (Fig. \ref{fig:breast2D}-C). We calculate the mean of the absolute value of the difference between the final reconstruction and the original model, which is displayed in Fig. \ref{fig:breast2D} with the symbol $\varepsilon$. We can see how the reconstruction closely matches the ground-truth model, both qualitatively and quantitatively. As expected, inaccuracies can be observed in the reconstruction, which can be explained through a number of factors. First, limited sampling of the wavefield is performed at the boundaries of the model because a finite number of receivers is used. Second, the available frequency bandwidth is also necessarily finite, which will limit resolution and prevent high-contrast interfaces from being perfectly recovered.

\begin{lstlisting}[language=Python3, float, floatplacement=H, label={lst:13}, caption={At every iteration, a subset of the available shots are selected randomly to calculate a gradient. The calculated gradient is then used to update the speed of sound distribution.}]
async def run_iter(f_max):
    # Select some shots for this iteration
    shot_ids = acquisitions.select_shot_ids(
                         num=shots_per_iter, 
                         randomly=True)

    # Clear the gradient
    vp.clear_grad()

    # Async loop over selected shots
    @runtime.async_for(shot_ids)
    async def loop(worker, shot_id):
        # Fetch one data point
        sub_problem = problem.sub_problem(
                                    shot_id)
        wavelets = sub_problem.shot.wavelets
        observed = sub_problem.shot.observed

        # Pre-process the wavelets
        wavelets = p_wavelets(wavelets,
                             runtime=worker)
        # Execute the PDE
        modelled = pde(wavelets, 
                       vp, 
                       problem=sub_problem, 
                       runtime=worker)

        # Pre-process traces
        traces = p_traces(modelled, 
                          observed, 
                          f_max=f_max,
                          runtime=worker)
        # Calculate loss
        fun = await loss(traces.outputs[0], 
                    traces.outputs[1],
                    problem=sub_problem, 
                    runtime=worker).result()

        # Calculate derivative
        await fun.adjoint()

    # Wait for loop to end 
    await loop
    # Update vp
    await optimiser.step()
\end{lstlisting}

\begin{lstlisting}[language=Python3, float, floatplacement=H, label={lst:14}, caption={The inversion is performed by selecting subsequent frequency bands and, in each band, a certain number of iterations are run to calculate a gradient.}]
opt_loop = OptimisationLoop()

# Start optimisation
for block, f_max in \ 
        opt_loop.blocks(num_blocks, freqs):
    # Every iteration in the block
    for iteration in \
            block.iterations(num_iters):
        await run_iter(f_max)
\end{lstlisting}

\begin{figure}[t]
    \centering
    \includegraphics[width=8.5cm]{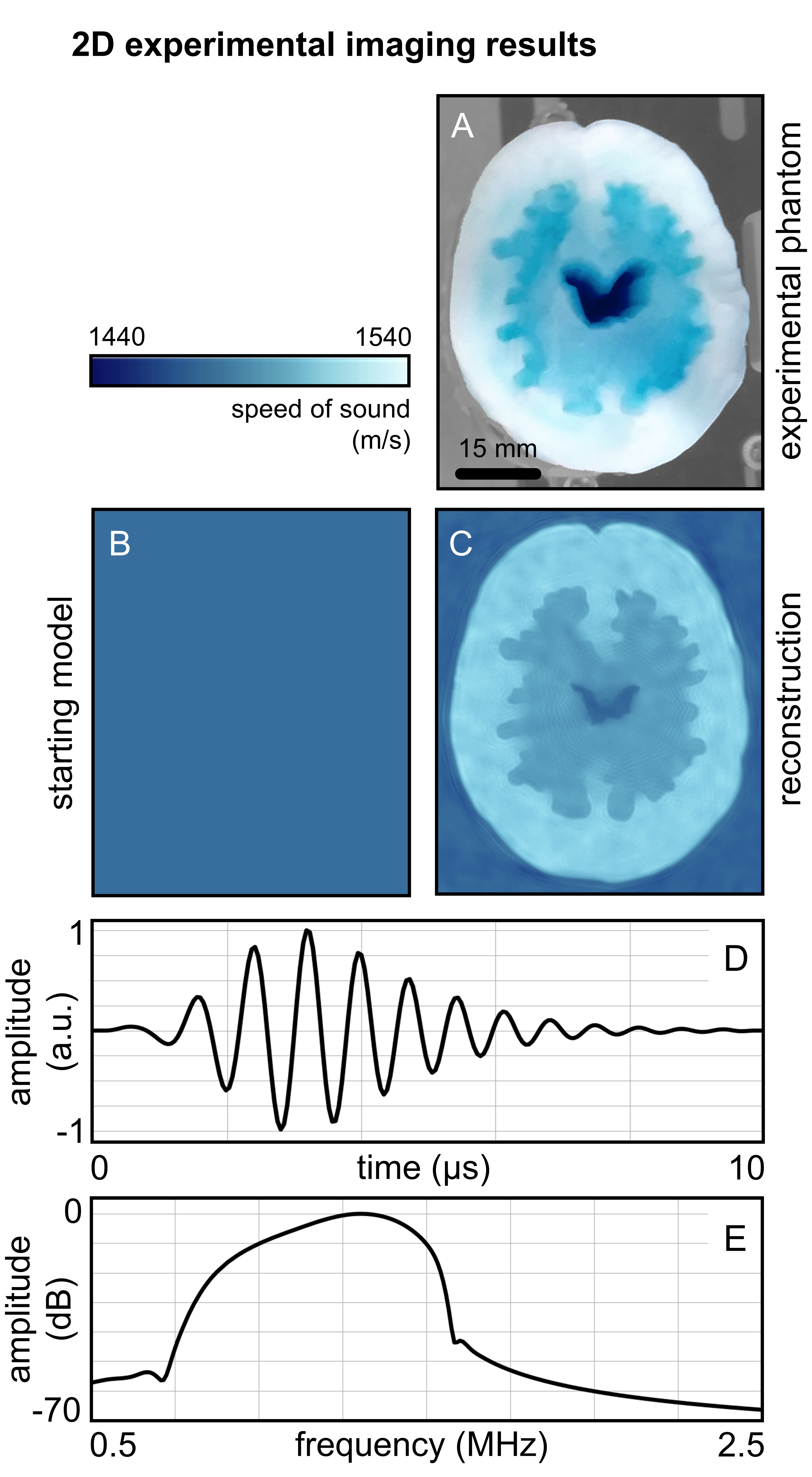}
    \caption{Experimental Stride reconstruction in 2D. A tissue-mimicking phantom (A) is imaged starting from a homogeneous distribution of speed of sound (B). The Stride reconstruction (C) closely matches the target phantom, is able to recover the different layers of speed of sound and the complex interface between those layers. We can also see the signal used experimentally for imaging (D) and its corresponding magnitude spectrum (E).}
    \label{fig:BB}
\end{figure}

Next, we apply the same imaging script that we have just introduced to now image an experimental tissue-mimicking phantom. A polyvinyl alcohol (PVA) cryogel phantom was constructed with two layers of different speed of sound values and an inner cavity filled with water \cite{Chee2016WalledDynamics}. The dimensions of the phantom are, approximately, 57.4 mm in width, 70.4 mm in height, and 130 mm in depth. Speed-of-sound values for the different layers of the phantom were experimentally measured using time of flight to be 1521 \textpm\ 3 m/s for the outer layer and 1502 \textpm\ 4 m/s for the inner layer. A photograph of the cross section of the phantom can can be seen in Fig. \ref{fig:BB}-A. Data were then acquired using two P4-1 transducers (ATL, USA), each of which contains 96 transmitting and receiving elements. The two P4-1 transducers were independently attached to two rotary motors, allowing them to move around the phantom for full illumination. Data were acquired by transmitting with a centre frequency of 1.4 MHz.


The inversion was performed over \ilu[120]{\micro\second}, in steps of \ilu[0.048]{\micro\second} (maximum CFL constant of 0.37), using a spatial sampling of 0.200 mm (minimum of 4.67 PPW) and a grid size of 890\texttimes 890. Imaging was carried out using a single block and a single frequency band with an upper limit of 700 kHz across a total of 152 iterations. During each iteration, 10 shots were selected randomly without replacement so that each shot was used four times at the end of the block. A single frequency band is sufficient in this example because, for this particular experiment, the starting point of our inversion is close enough to the minimum of the optimisation to ensure convergence. Simultaneously, the resolution offered by this frequency band (with a half-wavelength of approximately 1 mm in water) is sufficient, given the size and level of detail of the phantom, to recover a high-resolution reconstruction.

Using a starting model that contained homogeneous water (Fig. \ref{fig:BB}-B), a high-resolution reconstruction of the phantom is obtained (Fig. \ref{fig:BB}-C). Stride can successfully recover the two layers of speed of sound, as well as the internal water cavity. The reconstruction shows high contrast between layers, and the correct recovery of the complex details at the interface between them. We can also see how, at some points, the two layers of the phantom seem to gradually dissolve into one another instead of presenting sharp interfaces. This could be an imaging artefact due to errors in the calibration of the data acquisition setup, but could also be due to the natural degradation of the phantom, which could have led to the two layers merging at these locations.

We run this inversion on the same workstation as the previous example, using the same 3 Mosaic workers, so that each shot gradient calculation took a total of 28.32 \textpm\ 4.46 s. Adding a single argument to the PDE call, \texttt{pde(..., platform="nvidia-acc")}, is sufficient to run the same inversion on an available GPU instead of the CPU. In this case, the Devito-generated OpenACC solver is compiled using the PGI \texttt{pgc++} compiler version 21.2. Then, using the same workstation, equipped with an NVIDIA GeForce RTX 2080 Ti with 11 GB of memory, and a single Mosaic worker, each shot gradient calculation took a total of 5.63 \textpm\ 0.07 s.

\begin{figure*}[htp]
    \centering
    \includegraphics[width=17.5cm]{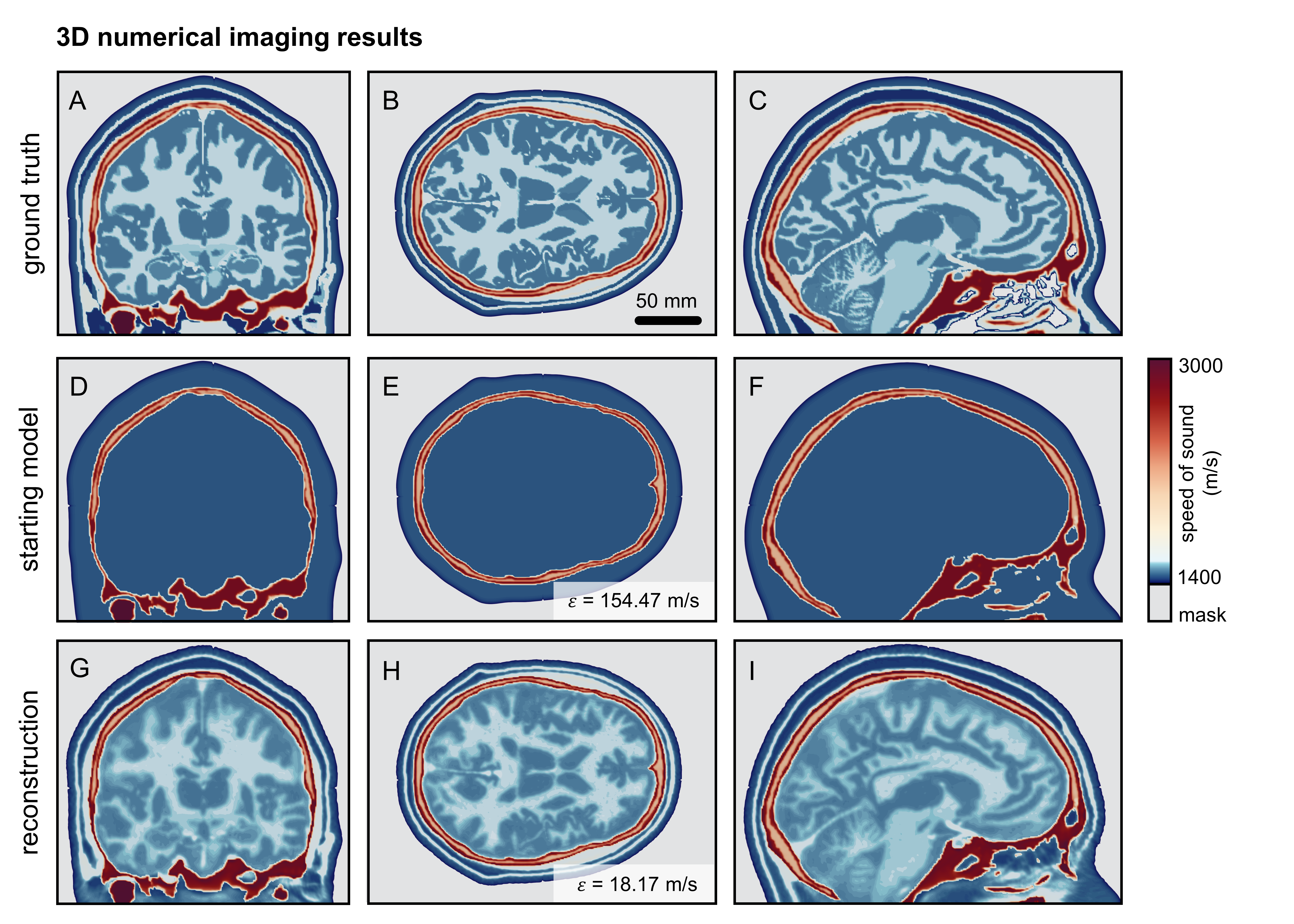}
    \caption{Stride reconstruction in 3D. A 3D acoustic head model (top row) is imaged starting from a model that contains only the skull and is homogeneous otherwise (middle row). Stride manages to accurately reconstruct the target model (bottom row). The mean of the absolute value of the difference between the ground-truth model and the inversion is displayed here as $\varepsilon$.}
    \label{fig:brain3D}
\end{figure*}

\begin{figure}[t]
    \centering
    \includegraphics[width=8.5cm]{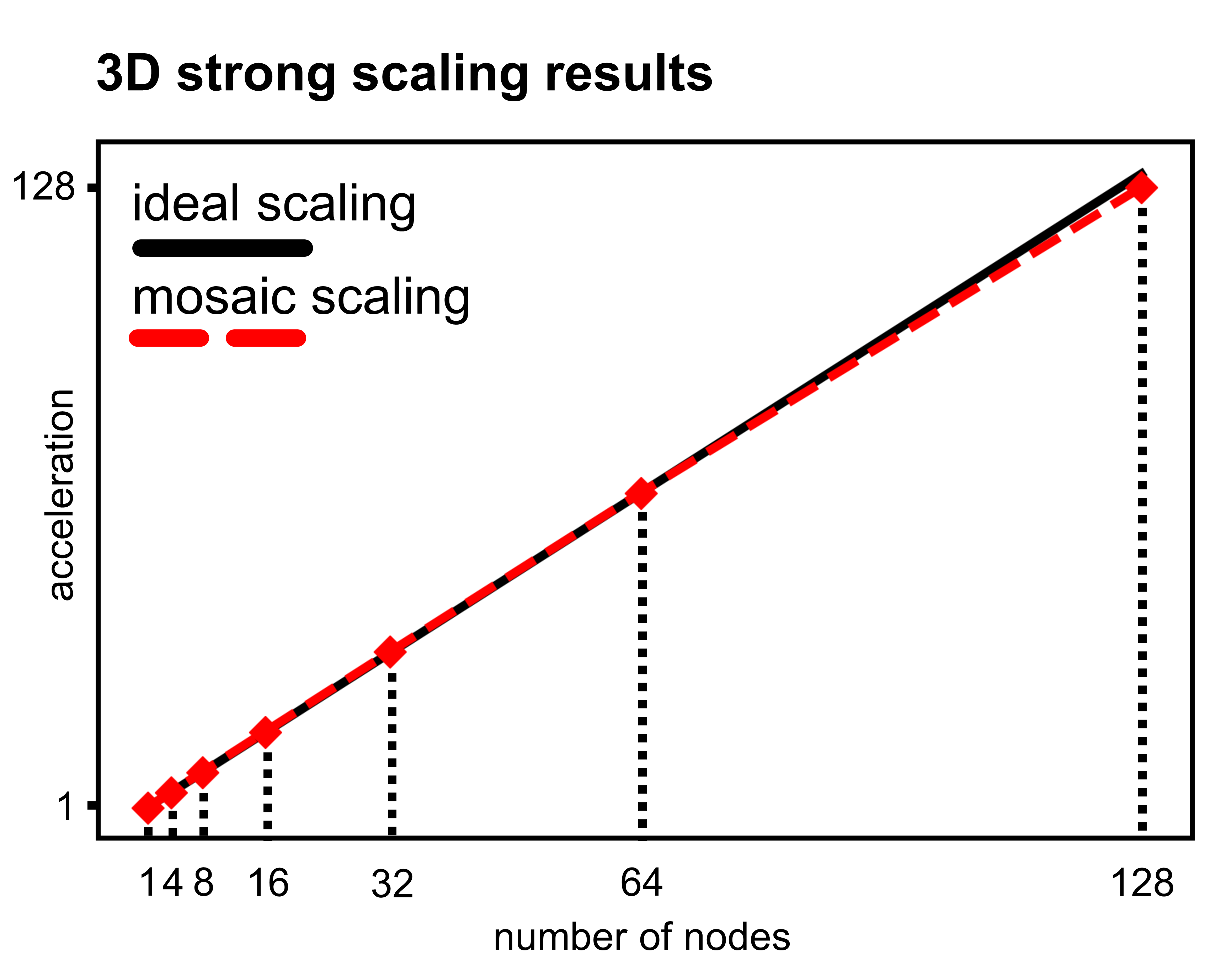}
    \caption{Mosaic strong scaling for the 3D head model. Scaling obtained with Mosaic (red, dashed line) is compared to the ideal scaling scenario (black, continuous line). Scaling is analysed by running 128 shot gradient calculations for the 3D head model across an increasing number of compute nodes. Acceleration is calculated as the amount of time taken to complete all gradient calculations using a certain number of nodes with respect to the time taken using a single node, averaged over 5 experiments.}
    \label{fig:scaling}
\end{figure}

\subsection{Imaging in 3D}

Although relevant when imaging structurally simple, soft tissues such as the breast, 2D imaging on its own is of limited applicability in realistic tomographic reconstructions, where 3D modelling and inversion is needed to account for the full physics of wave propagation in the human body. At the same time, it is in these 3D problems where the computational cost of FWI is most apparent and where tomography codes are required to scale robustly. In order to showcase the scaling capabilities of Stride, we choose for our second experiment a numerical 3D model of the adult human head (Fig. \ref{fig:geometry}-B). The model is based on the MIDA model \cite{Iacono2015MIDA:Neck}, to which acoustic properties were assigned as described by Guasch \textit{et al.} \cite{Guasch2020Full-waveformBrain}. Three slices through this numerical model can be seen in Fig. \ref{fig:brain3D}-A to C. The model is sampled with a spacing of 0.750 mm (minimum of 3.22 PPW), resulting in a grid of size 367\texttimes 411\texttimes 340 and more than 51 million unknown parameters to be estimated. A total of 1024 transducers were located around the head as seen in Fig. \ref{fig:geometry}-B, with all transducers acting both as sources and receivers. Imaging was performed with a three-cycle tone burst centred at 500 kHz. Modelling was carried out over \ilu[300]{\micro\second}, with time steps of \ilu[0.150]{\micro\second} (maximum CFL constant of 0.60).

Stride has been designed to seamlessly scale from 2D to 3D, and moving from one to the other only requires changing three lines of the code when defining the spatial grid. The remaining code can be run without any changes. In this case, the reconstruction is performed in the frequency range between 100 kHz and 600 kHz, starting from a model that only contains the skull (Fig. \ref{fig:brain3D}-D to F). Each frequency band in the reconstruction is run for 8 iterations, and 128 shots are randomly selected without replacement for each of them.

Due to the higher computational requirements in 3D, we run this reconstruction in an HPC cluster environment. Except for removing the \texttt{mosaic.interactive("on")} command, no changes are required to the code when scaling from the local to the cluster environment. Each compute node in the cluster is equipped with 256 GB of memory and 128 cores (2xAMD Zen2 EPYC 7742, 64 cores, 2.25 Ghz). Nodes are connected using an HPE Slingshot interconnect with 200 Gb/s signalling. The Devito solver is compiled using the GNU \texttt{gcc} compiler version 7.5, and is executed using OpenMP thread-level parallelism across 32 threads.

Each of the nodes calculates the gradient for a single shot at a time, which once more entails one forward propagation and one adjoint propagation of the acoustic solver, before combining the gradients for all shots at each iteration. Work distribution across the different nodes is managed by the Mosaic runtime, with the time taken to allocate this work generally dominated by the serialisation, communication, and processing of the data associated with the execution of each shot. However, serialisation in Mosaic has a negligible impact due to its zero-copy implementation. Communication overheads could have an impact on performance, but these are minimised by high-speed interconnects and by the asynchronous nature of Mosaic and its underlying ZeroMQ sockets. This means that user code is not slowed down by the actual time taken to send messages across the network by allowing the overlap of computation and communication: the \textit{head} process dispatches all shots almost instantaneously, and independent \textit{worker} processes across the network start computing as soon as the first message arrives. Message processing, on the contrary, will have an impact on performance due to the intrinsic single-threaded nature of Python. This could be alleviated by offloading some of this processing to lower-abstraction interfaces in C. With all this in mind, each shot gradient calculation took 5.82 \textpm\ 0.36 min, including time spent in work distribution.

The high accuracy of the final reconstruction obtained using Stride can be seen in Fig. \ref{fig:brain3D}-G to I. Also in this case, we have calculated a corresponding quantitative error measure for the full 3D model, shown in Fig. \ref{fig:brain3D} with the symbol $\varepsilon$. Errors in the reconstruction can in this case be attributed to similar reasons to the previous numerical 2D case, with the added factor of limited illumination in certain regions of the model. We can see, for example, how the regions close to the neck and around the sinuses are more poorly resolved due to the location of sources and receivers around the head. We can also see how resolution is degraded as we move closer to the upper regions of the skull due to lower ray density in these areas.

At this point, we explore the scaling capabilities of the Mosaic parallelisation layer by running a fixed number of individual shot gradient calculations, 128, while increasing the number of compute nodes used in the HPC cluster. The achieved acceleration is calculated by comparing the amount of time taken to complete all gradient calculations using a certain number of nodes with respect to the time taken using a single node. Under ideal circumstances, this means that, for example, an acceleration of 128 times is expected when using 128 compute nodes. This test is repeated five times, and the final acceleration is taken as the average over all repetitions.

Results for this strong scaling test can be seen in Fig. \ref{fig:scaling}, where we can observe that Mosaic achieves nearly ideal scaling up to 128 compute nodes. For the largest number of nodes, we can see how the obtained acceleration deviates slightly from the ideal curve. This corresponds, approximately, to a 2\% loss in performance, which can be attributed to the effective single-threaded nature of Python programs that we have previously discussed.

\section{Discussion}
\label{sec:discussion}

We have shown that Stride provides an intuitive framework for the solution of ultrasound tomography problems, seamlessly switching between 2D and 3D applications, and between a local workstation and a multi-node cluster.

Implementations of ultrasound tomography methods like FWI have to address their computational and algorithmic complexity. To do this, Stride has been designed to provide tailored optimisation routines, high-performance PDE solvers, and scalability to HPC systems, while simultaneously offering a high level of abstraction to ensure flexibility, productivity, and modularity.

From the point of view of the optimisation, we have seen how Stride closely matches the mathematical formulation of the inverse problem, for which gradients can be intuitively calculated using the adjoint method. Our approach here resembles that taken by machine learning libraries like PyTorch \cite{Paszke2017AutomaticPyTorch}, which have been highly successful at broadening the reach of these technologies beyond computational experts. This serves the double purpose of easing adoption by users, some of which might already be familiar with some of these libraries, and facilitating integration with these machine learning tools.

We have to note that gradients for Stride problems are calculated at a high level by treating the PDE or the loss functions as differentiable primitives, but no differentiation is happening through their internal mathematical operations. This is the subject of ongoing research and will be introduced in future versions of Stride.

From the point of view of the PDE solver, Stride faces the performance-flexibility dichotomy in a similar manner to the geophysical library JUDI \cite{Witte2019AJulia}: we provide intuitive interfaces in a high-abstraction language, while using a DSL like Devito under the hood. From a symbolic specification of the PDE, Devito automatically generates architecture-specific C code that matches the performance of hand-tuned implementations \cite{Louboutin2019DevitoExploration, Luporini2020ArchitectureComputation}. This offers a high degree of flexibility, allowing the inclusion of new physical models with minimal effort and without hindering performance. It is this flexibility that allows us to run the same wave equation solver on a CPU multi-threaded environment or a GPU with effectively no code changes.

Currently, Stride problems can only be defined on rectangular grids, on which finite-difference methods can be applied using Devito. Nonetheless, Stride does not prescribe any of these, and future work will explore the inclusion of different discretisation approaches and integration with other DSLs like FEniCS/Firedrake for finite-element methods \cite{Logg2012AutomatedEngineering,Rathgeber2016Firedrake:Abstractions} or Dedalus for spectral methods \cite{Burns2020Dedalus:Methods}.

Other open-source libraries exist for numerical modelling in ultrasound medical imaging, such as the previously mentioned kWave \cite{Treeby2010K-Wave:Fields}, based on pseudospectral element methods; Field II \cite{Jensen1996FIELD:Systems}, which uses a linear scattering approximation; or Bempp-cl \cite{Betcke2021Bempp-cl:Library.}, which employs a boundary element method, among others. These libraries have been tailored to accurately model sound propagation in biological tissues and generally provide hand-tuned implementations that can achieve high performance. Stride is agnostic to the underlying solver employed and any of these could be readily integrated with it. However, that would diminish the flexibility that is achieved by using a DSL that can obtain comparable performance for both the physical models currently available and any new ones that could be introduced.

Stride has been designed to tackle the problem of intuitively scaling to HPC systems in a similar spirit as for the solver: high-level interfaces hide from the user the complexity of deploying the algorithms to target systems, allowing imaging scientists to focus on the reconstruction algorithms rather than the low-level details. We provide for this the custom parallelisation library Mosaic.

Traditional HPC workloads usually rely on the message passing interface (MPI) standard to express parallelism in applications. However, originally designed in the 1990s, MPI has so far no capacity for fault tolerance and its interfaces are too cumbersome and low level for most non-specialists. Other Python libraries exist for writing parallel applications, most notably Dask \cite{DaskDevelopmentTeam2016Dask:Scheduling}, PyCOMPSs \cite{Tejedor2017PyCOMPSs:Python}, and Ray \cite{Moritz2017Ray:Applications}. Dask expresses parallelism as a series of stateless tasks that form a computational graph, which can be executed in parallel. PyCOMPSs uses tasks similarly to express parallelism, although these do not have to be stateless. However, PyCOMPSs employs a Java-based runtime that requires the serialisation of objects to file in order to communicate with Python, incurring in a performance penalty. Contrarily, the Ray parallel framework is primarily based on the actor model. We have chosen to design Mosaic using an actor-based model because, much like object-oriented programming, we consider that it better matches the world view and the mental framework of domain specialists. It also allows us to keep objects and their allocated memory warm within a specific compute node or associated accelerator, incidentally making it more intuitive for end users to manipulate remote memory. We have chosen to implement a custom parallelisation library for Stride due to a need for fine-grained control of the computational workload allocation and memory management that existing libraries are unable to provide.

Through the examples presented, we have seen that switching from a local multi-processing environment to an HPC cluster with Mosaic is straightforward and requires no significant code changes. We have also seen through our 3D experiments that realistic Stride reconstructions could be potentially scaled across hundreds of compute nodes thanks to the zero-copy, asynchronous work allocation of the Mosaic library. However, work is still needed to fully understand and exploit the scaling capabilities of Mosaic across large on-premises and cloud computing clusters, with particular interest in minimising data transfers across the network by exploiting caching mechanisms to detect redundant communications. 

Additionally, while Mosaic offers the capacity to parallelise across elements of an iteration batch, the integration with Devito offers another degree of freedom to parallelise within PDE solves through MPI-based domain decomposition. Domain decomposition, whose use in Stride is being actively explored, allows a user to distribute the computation of the PDE solution. This will be of importance when solving large problems whose size exceeds memory available in any single node or memory available in a particular accelerator such as a GPU. It will also allow for increased computational performance by splitting PDE solves in a single node across available CPU sockets, thus enforcing data locality.

There are two distinct applications for which Stride has been designed: wave propagation modelling and tomographic imaging. In terms of modelling wave phenomena, a number of other libraries are openly available to users, some of which include the already mentioned Field II \cite{Jensen1996FIELD:Systems}, Bempp-cl \cite{Betcke2021Bempp-cl:Library.}, or kWave \cite{Treeby2010K-Wave:Fields}, among others. The choice of one library over another will be down to the aims and requirements of a specific modelling exercise. For example, Field II should be chosen when modelling accuracy can be traded off for shorter computational times, whereas the boundary element method in Bempp-cl will provide accurate modelling that remains computationally efficient when the number of tissue interfaces in the model is low. As we have shown here, Stride and kWave can achieve similar levels of modelling accuracy for complex tissue geometries. Nonetheless, finite-difference solvers in Stride will be more computationally efficient, whereas kWave will display smaller numerical dispersion for a similar discretisation grid thanks to its pseudospectral formulation. These differences will, however, become irrelevant as other numerical methods are integrated into Stride: a different method will be chosen depending on the specific application.

In terms of tomographic imaging, it is important to distinguish between full-wave methods, such as FWI, and others, such as time-of-flight tomography and diffraction tomography. Stride is, at the time of this writing, the only openly available library for full-wave tomographic imaging in the medical context. Stride, however, does not currently provide solvers for other types of ultrasound tomography and other tools should be used in these cases \cite{Ali2019Open-sourceTomography}.

In terms of compatibility, Stride can be installed on Unix operating systems, and is compatible with Windows through the Windows Subsystem for Linux and through Docker containers.

Through these design decisions, Stride achieves flexibility and modularity, allowing each of its components to be modified independently or entirely substituted. At the same time, importance has been placed on ensuring that lower-level interfaces can be used to provide users with increasingly fine-grained control over the problem and its execution. Although we have designed Stride with ultrasound tomography in mind, the formulation of the physics-constrained optimisation problem is related to other imaging techniques, like optoacoustic tomography, and even calibration methods like spatial response identification. This makes Stride readily applicable to a number of medical ultrasound problems.

\section{Conclusions}
\label{sec:conclusions}

Advances in ultrasound-based imaging methodologies such as ultrasound computed tomography and optoacoustic tomography rely on increasingly complex mathematical and computational models. This puts a strain on researchers to both develop novel imaging algorithms and translate them into high-performance and scalable code, thus slowing scientific progress.

To bridge the gap between flexible development and real-life application, we have designed and developed Stride, an open-source Python library that is both intuitive and efficient. Stride allows algorithms to be written for a 2D model and be easily scaled up to 3D, and allows code to be tested on a local workstation and readily deployed to an HPC cluster. We achieve this by combining modular interfaces written in a high-abstraction language with automatically-generated, high-performance solvers, and with tailored parallelisation routines.

By providing high-level interfaces that intuitively match the representation of problems posed by domain specialists, and which are efficient and scalable out of the box, Stride has the potential to dramatically increase the productivity of imaging researchers. This will have a significant impact by accelerating the development of new ultrasound-based imaging technology and its translation from bench to bedside. Furthermore, other imaging applications where the efficient solution of physics-constrained optimisation problems is needed could also benefit from the general abstractions provided by Stride, such as non-destructive testing, aeronautics, or experimental fluid mechanics.

\section*{Declaration of competing interest}

The authors declare no competing interests.

\section*{Acknowledgements}

This work was supported by the Wellcome Trust [grant number 219624/Z/19/Z]. The work of Carlos Cueto was supported by the Engineering and Physical Sciences Research Council Centre for Doctoral Training in Medical Imaging [grant number EP/L015226/1]. The work of Oscar Bates was supported by the Engineering and Physical Sciences Research Council Centre for Doctoral Training in Neurotechnology [grant number EP/L016737/1]. This work used the ARCHER2 UK National Supercomputing Service. The authors would like to acknowledge the ITRUSST (International Transcranial Ultrasonic Stimulation Safety and Standards) planning group who developed the benchmark example used in Sec. \ref{sec:accuracy} and provided the kWave simulation results.

\bibliographystyle{cas-model2-names}

\bibliography{main.bib}

\end{document}